\documentclass[
superscriptaddress,
amsmath,amssymb,
aps,
floatfix,
english,
twocolumn
]{revtex4-2}

\usepackage{placeins}

\usepackage{import}
\usepackage{bibunits}
\defaultbibliography{references}
\defaultbibliographystyle{apsrev4-2}

\usepackage{graphicx}
\usepackage{svg}
\usepackage{dcolumn}
\usepackage{bm}
\usepackage{siunitx}
\usepackage[T1]{fontenc}
\usepackage[utf8]{inputenc}

\usepackage[unicode=true, colorlinks=true, citecolor={blue!80!black}, urlcolor={blue!50!black}, linkcolor = {blue!80!black}]{hyperref}

\newcommand{\ghz}{\:\mathrm{GHz}}

\newcommand{\invs}{\:\mathrm{sec^{-1}}}
\newcommand{\mk}{\:\mathrm{mK}}

\graphicspath{{Figures/}}

\begin{document}
\title{Coexistence of nonequilibrium density and equilibrium energy distribution of quasiparticles in a superconducting qubit}
\author{Thomas~Connolly}
\thanks{These two authors contributed equally.\\
tom.connolly@yale.edu, pavel.kurilovich@yale.edu}
\affiliation{Departments of Applied Physics and Physics, Yale University, New Haven, CT 06520, USA}
\author{Pavel~D.~Kurilovich}
\thanks{These two authors contributed equally.\\
tom.connolly@yale.edu, pavel.kurilovich@yale.edu}
\affiliation{Departments of Applied Physics and Physics, Yale University, New Haven, CT 06520, USA}
\author{Spencer~Diamond}
\affiliation{Departments of Applied Physics and Physics, Yale University, New Haven, CT 06520, USA}
\author{Heekun~Nho}
\affiliation{Departments of Applied Physics and Physics, Yale University, New Haven, CT 06520, USA}
\author{Charlotte~G.~L.~B\o ttcher}
\affiliation{Departments of Applied Physics and Physics, Yale University, New Haven, CT 06520, USA}
\author{Leonid~I.~Glazman}
\affiliation{Departments of Applied Physics and Physics, Yale University, New Haven, CT 06520, USA}
\author{Valla~Fatemi}
\affiliation{Departments of Applied Physics and Physics, Yale University, New Haven, CT 06520, USA}
\affiliation{School of Applied and Engineering Physics, Cornell University, Ithaca, NY 14853}
\author{Michel~H.~Devoret}\thanks{michel.devoret@yale.edu}
\affiliation{Departments of Applied Physics and Physics, Yale University, New Haven, CT 06520, USA}

\begin{abstract}

The density  of quasiparticles typically observed in superconducting qubits exceeds the value expected in equilibrium by many orders of magnitude. Can this out-of-equilibrium quasiparticle density still possess an energy distribution in equilibrium with the phonon bath? Here, we answer this question affirmatively by measuring the thermal activation of charge-parity switching in a transmon qubit with a difference in superconducting gap on the two sides of the Josephson junction. We then demonstrate how the gap asymmetry of the device can be exploited to manipulate its parity.
\end{abstract}

\maketitle

In thermal equilibrium, the breaking of Cooper pairs in conventional superconductors should be negligible when the temperature is much smaller than the superconducting gap. 
Experimentally, however, a finite fraction of broken Cooper pairs,
typically in the range of $10^{-9}-10^{-5}$, persists to the lowest measured temperatures. While it is appreciated that the resulting “resident” quasiparticles (QPs) are detrimental for superconducting devices \cite{glazman_bogoliubov_2021, aumentado_nonequilibrium_2004, lutchyn_quasiparticle_2005, lutchyn_kinetics_2006, shaw_kinetics_2008, martinis_energy_2009, de_visser_number_2011, catelani_relaxation_2011, catelani_quasiparticle_2011, catelani_decoherence_2012, de_visser_microwave-induced_2012,
levenson-falk_single-quasiparticle_2014,
pop_coherent_2014, wang_measurement_2014, yan_flux_2016,grunhaupt_loss_2018, kurter_quasiparticle_2022}, their origin is not fully understood. {The main conjectures are that QPs might be generated by stray mm-wave photons, radioactive materials, or cosmic rays \cite{bespalov_theoretical_2016, vepsalainen_impact_2020, martinis_saving_2021, wilen_correlated_2021, cardani_reducing_2021, diamond_distinguishing_2022, mcewen_resolving_2022, mannila_superconductor_2022, iaia_phonon_2022, bargerbos_mitigation_2023}.}

{Another long-standing mystery is the energy distribution of the resident QPs. Can the QP \textit{energy} distribution be in thermal equilibrium with the phonon bath, despite their \textit{density} being out-of-equilibrium?
{Previous experiments investigated this question by measuring $1e$ switching in the offset charge of a transmon qubit. These offset-charge parity switches were identified with the tunneling of resident QPs \cite{riste_millisecond_2013, sun_measurements_2012, serniak_hot_2018}. The authors found that the charge-parity switches are approximately equally likely to excite or relax the qubit, leading them to the conclusion that the QP energy distribution is ``hot''. Later, an alternative explanation for the results of \cite{serniak_hot_2018} emerged, which does not require the resident QPs to be hot \cite{houzet_photon-assisted_2019}. In this alternative mechanism, no pre-existing QPs in the device are required to cause a parity switch. Instead, a stray photon with energy greater than $2\Delta$ is absorbed at the Josephson junction. This process \textit{simultaneously} breaks a Cooper pair and deposits one of the resulting quasiparticles on each side of the junction, switching the charge-parity [see Fig.~\ref{fig:1}(b)]. Further experiments \cite{diamond_distinguishing_2022} confirmed the prevalence of the photon-assisted parity switching mechanism in \cite{serniak_hot_2018}. The energy distribution of the resident QPs in superconducting qubits was thus mischaracterized. To reveal it, photon-assisted parity switching must be suppressed.}

\begin{figure}[t]
\includegraphics[scale=1.0]{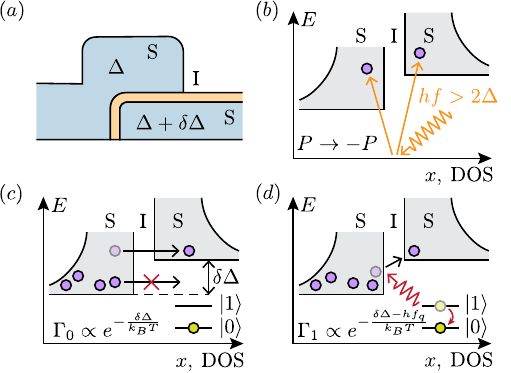}
\caption{
(a) Schematic of the junction between two superconducting films with gaps {$\Delta$ and $\Delta + \delta\Delta$}.
The gap difference, $\delta\Delta$, arises due to the thickness difference of the films \cite{chubov_dependence_1969, yamamoto_parity_2006, court_energy_2008}.
{(b) Photon-assisted parity switching. A stray photon breaks a Cooper pair at the junction, depositing one QP on each of its electrodes, and flipping the parity. This process overwhelmed the tunneling of resident QPs in Ref.~\cite{serniak_hot_2018}.}
(c) At small temperatures, the QP density is much higher than the value expected in thermal equilibrium, but the QP distribution function is thermalized to the temperature $T$ of the phonon bath.
{As a result, when $\delta\Delta \gg k_B T$, the QP tunneling rate activates with temperature.}
(d) In the excited state, the qubit energy, $h f_q$, can be transferred to a QP, helping it traverse the gap difference.
This reduces the activation energy by $h f_q$.
\label{fig:1}}
\end{figure}


{Here, we suppress the rate of parity switching by stray photons to $\Gamma_0 = 0.14 \pm 0.01\invs$ in a 3D aluminum transmon, an improvement of three orders of magnitude compared to a previous measurement of the same device \cite{serniak_direct_2019}. {We achieve this by enhancing the filtering and shielding against high-frequency photons. With enhanced filtering, parity switching becomes dominated by the tunneling of resident QPs.}

We observe Arrhenius activation of the parity switching rates with activation energies much smaller than the superconducting gap, which suggests that the QP energy distribution is indeed thermalized to the phonon bath, despite the non-equilibrium QP density. We attribute the activation to the difference of superconducting gaps, $\delta\Delta$, between the two sides of the junction impeding QP tunneling [see Fig.~\ref{fig:1}].}

{The activation energies for QP tunneling are very different for the ground and the excited state of the transmon. For the ground state, the activation energy is given by $\delta\Delta$ [see Fig.~\ref{fig:1}(c)]. When the transmon is excited, the QP can absorb the energy from the qubit when tunneling across the junction [see Fig.~\ref{fig:1}(d)]. This reduces the activation energy to $\delta\Delta - h f_q$, where $f_q$ is the qubit frequency. Thus, the parity switching rate when the qubit is excited vastly exceeds the ground state rate at low temperatures. We exploit this asymmetry to control the steady-state charge-parity of the device.}

\begin{figure}[t]
    \includegraphics[scale=1]{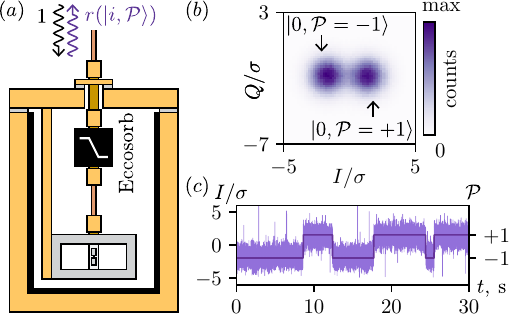}
    \caption{
    Measurement of the parity switching rate in the transmon ground state. (a) Schematic of the measurement setup. 3D transmon \cite{koch_charge-insensitive_2007, paik_observation_2011} is placed in a light-tight shield. (b) Histogram of single-shot cavity measurements. Two distributions correspond to the ground state of the transmon in the two parity states ($\mathcal{P}=\pm1$). (c) Measurement record of the $I$ quadrature over a 30 second time interval. Solid line shows a hidden Markov model state assignment. \label{fig:2}}
\end{figure}

We begin by noting that monitoring of the transmon parity is allowed by the fact that parity switching effectively shifts the offset charge by $1e$. Then, if the transmon spectrum is sensitive to the offset charge, parity switching events can be detected. To attain the offset charge sensitivity, we use a transmon with a moderate ratio of Josephson and charging energies, $E_J/E_C = 17.5$. In our system, this makes the frequency pull of the resonator different for the four relevant states constituting both charge parities for the $|0\rangle$ and $|1\rangle$ transmon states. This, along with a relatively long energy relaxation time $T_{1} = 193 \pm 25 \:\mathrm{\mu s}$, and the use of a quantum-limited amplifier \cite{frattini_optimizing_2018}}, allows us to distinguish these four states with a single-shot readout \cite{serniak_direct_2019} at certain values of the offset charge. For the data presented in the experiment, we use $n_g = 0.163\pm 0.003$ (in units of $2e$). At this value of offset charge, the dephasing time is $T_2^\star = 3 \pm 1\:\mathrm{\mu s}$

To suppress the flux of high-frequency photons incident on the transmon, the device is placed in a shield [see Fig.~\ref{fig:2}(a)] with seams sealed by indium o-rings and interior walls coated with an absorptive carbon-impregnated epoxy. A dissipative Eccosorb CR-110 low pass filter is placed on the microwave line inside the shield. At the minimum Cooper pair breaking frequency of $2\Delta / h \sim 100\ghz$, we expect \cite{halpern_far_1986, danilin_engineering_2022} this filter to provide $35\:\mathrm{dB}$ of attenuation \footnote{The disadvantage of our filter is $\sim 13\:\mathrm{dB}$ of insertion loss at the readout frequency.
This required a long $11\:\mu\mathrm{s}$ readout pulse to achieve a single-shot readout, which is still possible because $T_1 = 193 \pm 25\:\mathrm{\mu s}$ for our qubit.}. {With this improved filtering, we observe much smaller parity switching rates and QP density compared to previous experiments \cite{serniak_hot_2018, serniak_direct_2019, diamond_distinguishing_2022}. More recent experiments have also reduced the flux of pair-breaking photons by engineering the modes of the circuit above $f = 2\Delta / h$ \cite{pan_engineering_2022,liu_quasiparticle_2022}.

Next, we measure the parity switching rate of the qubit by collecting a series of jump traces with one measurement of the qubit state every $2\:\mathrm{ms}$ \cite{noauthor_see_nodate}. A histogram of measurement outcomes is shown in Fig.~\ref{fig:2}(b) and a time trace is shown in Fig.~\ref{fig:2}(c). The two visible distributions correspond to the two charge-parities of the transmon in the ground state.
By fitting the jump traces to a hidden Markov model \cite{schreiber_pomegranate_2018}, we extract a parity switching rate of $\Gamma_0 = 0.14 \pm 0.01\invs$ \cite{noauthor_see_nodate}. Since the residual excited state population in the described measurement is small, $\lesssim 0.2\%$, the measured $\Gamma_0$ well approximates the parity switching rate in the ground state of the transmon. We note that, in contrast to previous experiments \cite{serniak_hot_2018, serniak_direct_2019}, our excited state population is not limited by photon-assisted processes.

\begin{figure}[t]
    \includegraphics[scale=0.977]{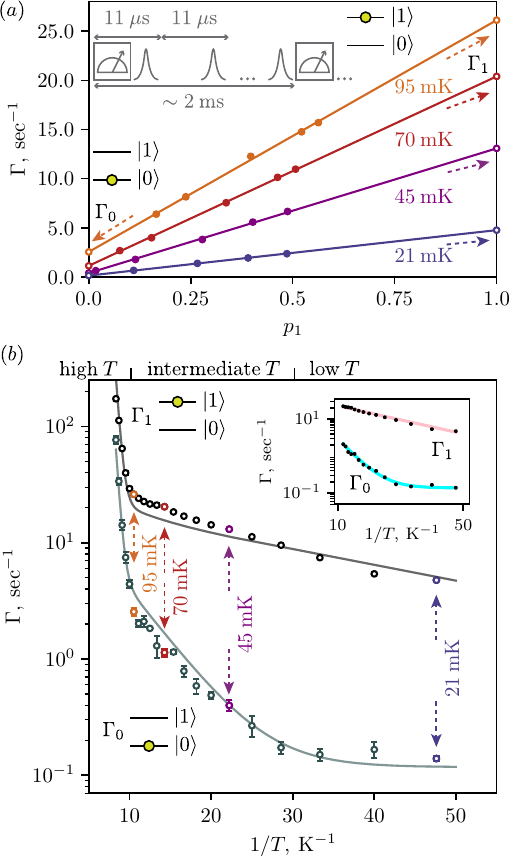}
    \caption{
    Temperature dependence of the parity switching rates in the ground and excited transmon states ($\Gamma_0$ and $\Gamma_1$, respectively). {Unless indicated with error bars, errors are within the size of the points.}
    (a) Parity switching rate as a function of excited state population $p_1$. Extrapolation to $p_1 = 0$ and $p_1 = 1$ yields $\Gamma_0$ and $\Gamma_1$. Different colors correspond to different mixing chamber temperatures. 
    (b) Temperature dependence of $\Gamma_0$ and $\Gamma_1$. Solid lines show the result of a simultaneous fit of our theory to both rates (see details in the text). {Inset shows the independent fit of $\Gamma_0$ and $\Gamma_1$ to an Arrhenius law in the intermediate temperature regime. In the fits a small offset of $0.14\:\mathrm{sec}^{-1}$ is added to the rates to account for the saturation of $\Gamma_0$. The difference in extracted activation energies between $\Gamma_0$ and $\Gamma_1$, $\delta E_A / h = 3.4 \pm 0.2\:\mathrm{GHz}$, is close to the qubit frequency, $f_q = 3.826\:\mathrm{GHz}$.}
    \label{fig:3}}
\end{figure}

By itself, $\Gamma_0$ is not enough to distinguish the contributions of non-equilibrium QP tunneling and photon-induced Cooper pair breaking. Therefore, we also measure the parity switching rate in the excited state of the transmon, $\Gamma_1$. 
We do this by interleaving readout pulses played every $2\:\mathrm{ms}$ with variable-amplitude qubit-scrambling pulses played every $11\:\mu\mathrm{s}$ \cite{diamond_distinguishing_2022}. The Gaussian qubit-scrambling pulses have a standard deviation of $10\:\mathrm{ns}$ and lead to a steady-state excited state population that we measure simultaneously with the parity.
The resulting dependence of the parity switching rate on the excited state population $p_1$ is well-described by $\Gamma = (1-p_1)\Gamma_0 + p_1 \Gamma_1$ [see Fig.~\ref{fig:3}(a)]. 
Extrapolating this line to $p_1=0$ and $p_1=1$ yields $\Gamma_0$ and $\Gamma_1$.
We obtain $\Gamma_0 = 0.14 \pm 0.01\invs$ and $\Gamma_1 = 4.76\pm0.04\invs$.
Remarkably, $\Gamma_1$ is approximately $34$ times higher than $\Gamma_0$. 
This observation is inconsistent with parity switching due to photon-induced pair-breaking, which should lead to $\Gamma_0 \sim \Gamma_1$ \cite{serniak_hot_2018, houzet_photon-assisted_2019, serniak_direct_2019, glazman_bogoliubov_2021, diamond_distinguishing_2022}. 
The remaining explanation is that resident QPs dominate parity switching in the excited state.

Next, to shed light on the energy distribution of the QPs, we measure $\Gamma_0$ and $\Gamma_1$ at different temperatures [see Fig.~\ref{fig:3}(b)].
The temperature dependence is distinct in three temperature regions, marked by ticks at the top of Fig.~\ref{fig:3}(b). In the high-temperature region, $T\gtrsim100\:\mathrm{mK}$, both rates rapidly activate with activation energies $\sim h\times50\:\mathrm{GHz}$. In the intermediate-temperature region, $30\:\mathrm{mK}\lesssim T\lesssim100\:\mathrm{mK}$, rates $\Gamma_0$ and $\Gamma_1$ also thermally activate, but with much smaller activation energies. Notably, the activation energies are very different for the two qubit states. Finally, in the low-temperature region, $T\lesssim 30\:\mathrm{mK}$, $\Gamma_0$ saturates to a constant value, while $\Gamma_1$ continues to activate.

We attribute the temperature dependence of the rates to QP tunneling in the presence of gap difference at the junction. Within this model, the parity switching rates are proportional to the density of the QPs. In the low gap film of the device, we model the QP density (normalized by the density of Cooper pairs) as
{\begin{equation}
\label{eq:qps}
x_\mathrm{qp} = x_\mathrm{qp}^\mathrm{res} + \sqrt{\frac{2\pi k_B T}{\Delta}} \exp\left(-\frac{\Delta}{k_B T}\right).
\end{equation}
The} second term describes the density of QPs in thermal equilibrium. Thermal QPs dominate $x_\mathrm{qp}$ in the high-temperature region of Fig.~\ref{fig:3}, $T\gtrsim 100\:\mathrm{mK}$, thus explaining high activation energy there. On the contrary, at $T\lesssim 100\: \mathrm{mK}$, resident QPs with a temperature-independent density $x_\mathrm{qp}^\mathrm{res}$ dominate $x_\mathrm{qp}$.

To describe the data at intermediate temperatures, we calculate the QP tunneling rates in a model with different gaps at the two sides of the junction. We assume that the energy distribution of the resident QPs is in equilibrium with the phonon bath. The presence of large low-gap pads ($\Delta$) on both sides of the junction {(see Fig.~S6 in the supplementary materials \cite{noauthor_see_nodate})} dictates the form of the distribution function $\mathcal{F}(\epsilon) = x_\mathrm{qp} \sqrt{\frac{\Delta}{2\pi k_B T}} e^{- \frac{\epsilon - \Delta}{k_B T}}$ for either side at sufficiently low temperatures ($k_B T \ll \delta\Delta$). This means that the QPs in both pads primarily reside in the low-gap regions. Then, when the qubit is in the ground state, the QP tunneling rate reads
\begin{equation}
    \label{eq:1}
    \Gamma_0^{\mathrm{qp}} = \eta f_q x_{\mathrm{qp}} \exp\left({-\frac{\delta\Delta}{k_B T}}\right), 
\end{equation}
where $\eta=4\sqrt{\frac{E_J}{E_C} \frac{\delta\Delta}{\Delta}} + \sqrt{\frac{2\Delta}{\delta\Delta - h f_q}}$ is a dimensionless pre-factor. Here, we assumed $\delta\Delta \ll \Delta$ and $k_B T \ll \delta\Delta - h f_q$ for simplicity. To explain the origin of the activation law in Eq.~\eqref{eq:1}, we note that the tunneling of QPs is suppressed while the transmon is in the ground state, since the QPs do not have enough energy to cross to the high-gap side of the junction [see Fig.~\ref{fig:1}(c)]. As the temperature is increased, a growing number of QPs have sufficient energy to tunnel, resulting in the activation of parity switching with activation energy $\delta\Delta$.
This explains the behavior of $\Gamma_0$ observed in the  intermediate temperature regime of Fig.~\ref{fig:3}.

When the qubit is in the excited state, the qubit energy, $h f_q$, can be transferred to tunneling QPs, helping them overcome the gap difference [see Fig.~\ref{fig:1}(d)]. 
The resulting activation energy for $\Gamma_1$ becomes $\delta\Delta - h f_q$. Explicitly, we find
\begin{equation}
    \label{eq:3}
    \Gamma_1^{\mathrm{qp}} = f_q \sqrt{\frac{2\Delta}{\delta\Delta - h f_q}} x_{\mathrm{qp}} \exp\left(-\frac{\delta\Delta - h f_q}{k_B T}\right).
\end{equation}
Here, we again assumed that $\delta\Delta - h f_q \gg k_B T$ and $\delta\Delta \ll \Delta$. 
The difference in activation energies for $\Gamma_1$ and $\Gamma_0$ leads to large asymmetry between these rates. Eq.~\eqref{eq:3} explains the behavior of $\Gamma_1$ in both the intermediate and the small temperature regimes.

In the low-temperature region of Fig.~\ref{fig:3}, $\Gamma_0$ saturates. We attribute this to photon-assisted processes \cite{noauthor_see_nodate} because the continued activation of $\Gamma_1$ in this regime rules out elevated temperature of the resident QPs. The corresponding contributions to the parity switching rates are temperature-independent and are in addition to $\Gamma_0^\mathrm{qp}$ and $\Gamma_1^\mathrm{qp}$ given by Eqs.~\eqref{eq:1} and \eqref{eq:3}. The contribution of photon-assisted processes to $\Gamma_1$ is negligible compared to QP tunneling.

Our model quantitatively matches the data. The joint fit to $\Gamma_0$ and $\Gamma_1$, shown as solid lines in Fig.~\ref{fig:3}(b), includes an extended version of the theory described in  Eqs.~\eqref{eq:qps}-\eqref{eq:3} that does not require $\delta\Delta\ll \Delta$ or $\delta\Delta - h f_q \gg k_B T$ \cite{noauthor_see_nodate}. From the joint fit we extract $\delta\Delta/h = 4.52\ghz$, $\Delta/h = 46\ghz$, $x_\mathrm{qp}^\mathrm{res} = 5.6 \times 10^{-10}$.
The extracted value of $\delta\Delta$ is close to that measured by a different method in a flux-tunable transmon of a similar design \cite{diamond_distinguishing_2022}. {The extracted values of $\Delta$ and $\delta\Delta$ are also close to the ones expected based on the thickness of the films used in our transmon, $20\:\mathrm{nm}$ and $30\:\mathrm{nm}$ \cite{chubov_dependence_1969, yamamoto_parity_2006, court_energy_2008}.}
To our knowledge, our observed $x_\mathrm{qp}^{\mathrm{res}}$  is the smallest reported to-date, which we attribute to improved filtering and shielding. {Some discrepancy between the model and the data seen in Fig.~\ref{fig:3} might arise due to the non-uniformity of superconducting gap in the films.}

In addition to jointly fitting all data to our full model, we independently fit $\Gamma_0$ and $\Gamma_1$ to activation laws in the intermediate and small temperature regimes, as shown in the inset of Fig.~\ref{fig:3}(b). We find a difference in activation energies of $\delta E_A / h = 3.4 \pm 0.2 \:\mathrm{GHz}$, close to the independently-measured qubit frequency $f_q = 3.826 \: \mathrm{GHz}$.

\begin{figure}[t]
\includegraphics[scale=1.0]{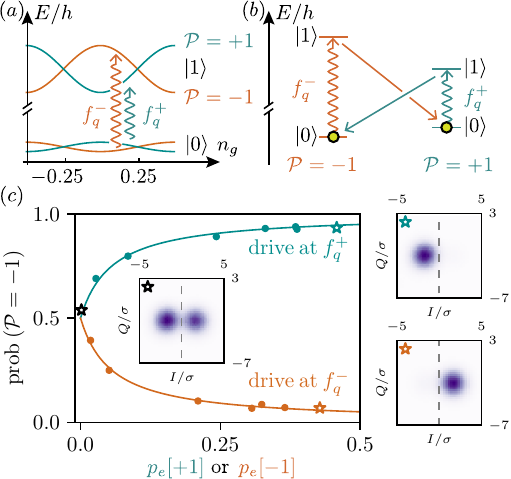}
\caption{
Manipulation of the transmon charge-parity. {The measurement is carried out at $T = 21\:\mathrm{mK}$ and $n_g = 0.163$ (in units of $2e$).} The pulse sequence is the same as in Fig.~\ref{fig:3}(a), except we use parity-selective $400 \:\mathrm{ns}$ scrambling pulses.
(a) Sketch of transmon energy as a function of the offset charge for $|0\rangle$ and $|1\rangle$ states with different parities, $\mathcal{P}=\pm 1$. 
Arrows depict the allowed transitions.
(b) Since the parity switching rate in $|1\rangle$ is large, selective driving of the qubit transition for one of the parities brings the system to the ground state with the opposite parity.
(c) Steady-state probability of $\mathcal{P}=-1$ as a function of the strength of driving applied at $f_q^+$ (teal) or $f_q^-$ (orange). As a proxy for drive strength, we use the probability of the qubit excited state $p_e$ conditioned on parity being $+1$ ($-1$).
{Solid lines are fit to a detailed balance model \cite{noauthor_see_nodate}.}
Insets show the measurement histograms in the absence of qubit driving and in the presence of parity-selective driving.
\label{fig:4}}
\end{figure}

Finally, we show that the parity of the transmon can be manipulated by applying a microwave drive. This manipulation is possible because (i) the transmon frequency $f_q^\mathcal{P}$ depends on the charge-parity $\mathcal{P} = \pm 1$, and (ii) the parity switching rate is much higher when the transmon is in the excited state ($\Gamma_1 \gg \Gamma_0$). {We note that manipulation of parity was recently reported for Andreev levels \cite{wesdorp_dynamical_2023, ackermann_dynamical_2023}.}

To explain the parity manipulation, we focus on the idealized case $\Gamma_0 = 0$. Then we show how to prepare the ground state with parity $-1$, assuming that initially either parity is equally probable. To bias the distribution towards $-1$, we drive the transmon at frequency $f_q^+$. The drive excites the transmon only if its ground state had parity $+1$. Subsequent relaxation may bring the transmon to the ground state with parity $-1$. The latter is a dark state of the drive and, therefore, the distribution becomes biased towards $-1$. Subsequent repetition of this process could fully polarize the system in the ground state with parity $-1$. Nonzero rate $\Gamma_0$ prevents full polarization, but close to unity polarization can be achieved as long as $\Gamma_1\gg \Gamma_0$. {The results of our measurements for how the degree of parity polarization depends on the strength of the driving are shown in Fig.~\ref{fig:4}.}

{Ability to manipulate parity can enhance qubit readout. In transmons, the readout drive has been found to cause undesired offset charge-dependent transitions to highly excited states \cite{sank_measurement-induced_2016, khezri_measurement-induced_2022}. Therefore, readout might be improved by choosing the transmon parity with the lowest rate of undesired transitions.}

In this Letter, we presented measurements of the charge-parity switching rates in the ground and the excited state of a transmon qubit. 
In our device, the latter rate is dominated by the tunneling of excess resident QPs.
We showed that the tunneling of non-equilibrium QPs activates with temperature [see Fig.~\ref{fig:3}].  
We interpret this as a consequence of superconducting gaps being different on the two sides of the Josephson junction [see Fig.~\ref{fig:1}]. 
Continued activation down to $T \lesssim 30\mk$ implies that the QPs are well-thermalized to the mixing chamber of our refrigerator, despite their out-of-equilibrium density.
Therefore, although the QPs may arise from high-energy sources, rapid inelastic processes \cite{kaplan_quasiparticle_1976, savich_quasiparticle_2017, catelani_non-equilibrium_2019} restore their near-thermal energy distribution. Our experimental results indicate that fabricating qubits with $\delta \Delta$ larger then $h f_q$ can suppress QP-induced decoherence \cite{marchegiani_quasiparticles_2022}. Gap difference in conjunction with fast QP energy relaxation may also reduce correlated errors \cite{wilen_correlated_2021, martinis_saving_2021, mcewen_resolving_2022} in quantum processors resulting from QP bursts produced by ionizing radiation \cite{vepsalainen_impact_2020, cardani_reducing_2021, diamond_distinguishing_2022}.

\emph{Acknowledgements.} ---  We acknowledge initial contributions of Kyle Serniak and Max Hays. We thank Vladislav D. Kurilovich, Manuel Houzet, Roman Lutchyn, and Rodrigo G. Corti\~nas for insightful discussions. We thank Gangqiang Liu, Vidul Joshi, and Maxime Malnou for providing the parametric amplifier. We thank Alessandro Miano for the help with the measurement setup. This research was sponsored by the Army Research Office (ARO) under grant numbers W911NF-18-1-0212, W911NF-22-1-0053, and W911NF-23-1-0051, by the Air Force Office of Scientific Research (AFOSR) under award number FA9550-19-1-0399, by the Office of Naval Research (ONR) under award number N00014-22-1-2764, and by the U.S. Department of  Energy, Office of Science, National Quantum Information Science Research Centers, Co-design Center for Quantum Advantage (C2QA) under contract number DE-SC0012704. The views and conclusions contained in this document are those of the authors and should not be interpreted as representing the official policies, either expressed or implied, of the U.S. Government. The U.S. Government is authorized to reproduce and distribute reprints for Government purposes notwithstanding any copyright notation herein.
\bibliography{references}
\end{document}


\myexternaldocument{ms}
\beginsupplement

\title{Supplementary information for ``Coexistence of nonequilibrium density and equilibrium energy distribution of quasiparticles in a superconducting qubit''}
\author{Thomas~Connolly}
\thanks{These two authors contributed equally.\\
tom.connolly@yale.edu, pavel.kurilovich@yale.edu}
\affiliation{Departments of Applied Physics and Physics, Yale University, New Haven, CT 06520, USA}
\author{Pavel~D.~Kurilovich}
\thanks{These two authors contributed equally.\\
tom.connolly@yale.edu, pavel.kurilovich@yale.edu}
\affiliation{Departments of Applied Physics and Physics, Yale University, New Haven, CT 06520, USA}
\author{Spencer~Diamond}
\affiliation{Departments of Applied Physics and Physics, Yale University, New Haven, CT 06520, USA}\author{Heekun~Nho}
\affiliation{Departments of Applied Physics and Physics, Yale University, New Haven, CT 06520, USA}
\author{Charlotte~G.~L.~B\o ttcher}
\affiliation{Departments of Applied Physics and Physics, Yale University, New Haven, CT 06520, USA}
\author{Leonid~I.~Glazman}
\affiliation{Departments of Applied Physics and Physics, Yale University, New Haven, CT 06520, USA}
\author{Valla Fatemi}
\affiliation{Departments of Applied Physics and Physics, Yale University, New Haven, CT 06520, USA}
\affiliation{School of Applied and Engineering Physics, Cornell University, Ithaca, NY 14853}
\author{Michel~H.~Devoret}\thanks{michel.devoret@yale.edu}
\affiliation{Departments of Applied Physics and Physics, Yale University, New Haven, CT 06520, USA}

\date{\today}

\maketitle

\tableofcontents

\vspace{2 cm}

\newpage

\section{Experimental setup}
\subsection{Diagram of the setup}
\begin{figure}[h]
\includegraphics[scale=1.0]{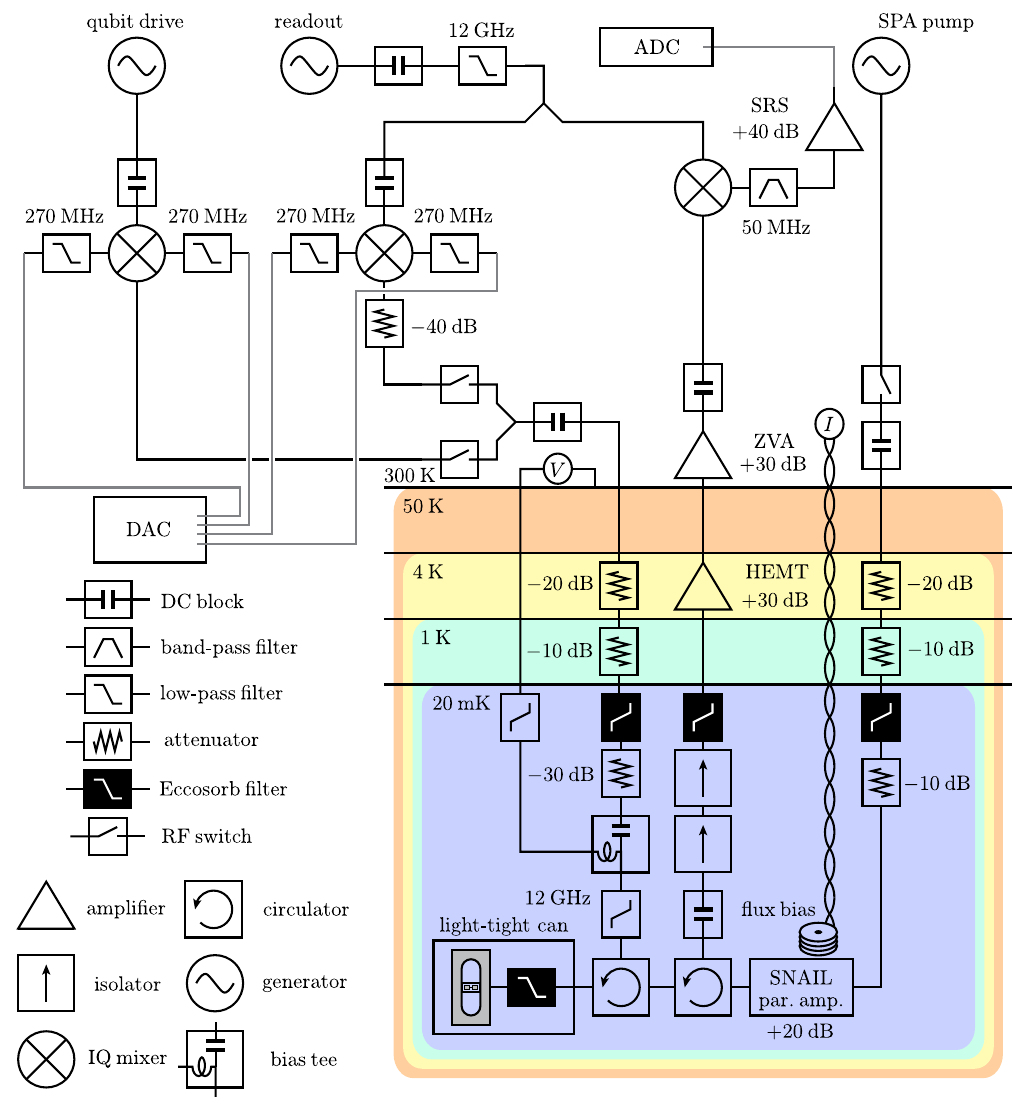}
\caption{Measurement setup. The oeration of a SNAIL parametric amplifier is described in detail in \cite{frattini_optimizing_2018}. Eccosorb low-pass filters are used to absorb radiation at the Cooper-pair breaking frequencies \cite{halpern_far_1986}, see Section~\ref{sec:filtering-configs} for a more in-depth discussion. An FPGA-based DAC is used to generate qubit pulses, readout pulses, and RF switch triggers; the ADC is used to analyze the readout outcomes. The bias tee is used to control the offset charge by applying a voltage to the resonator coupling pin.
\label{fig:wiring}}
\end{figure}

\subsection{Comparison of the filtering configurations}
\label{sec:filtering-configs}
In order to study the tunneling of quasiparticles (QPs) already present in the device, we must first reduce the flux of high-frequency photons which induce parity-switching events and generate additional QPs. Previous work \cite{serniak_direct_2019} has suggested that both photons propagating in the microwave lines and photons that leak through seams in the device packaging can be significant sources of parity switching \cite{diamond_distinguishing_2022}. To protect our sample from these photons, we iterated over several configurations of microwave line filtering and device shielding.

\subsubsection{In-line filtering}
To protect the device from pair-breaking photons present in the microwave lines, we use low-pass filters constructed by filling a segment of coaxial transmission line with Eccosorb CR-110, a lossy magnetically loaded epoxy. {We use Eccosorb since it has known high-frequency attenuation at 4 Kelvin \cite{halpern_far_1986}.
The attenuation coefficient of Eccosorb} increases from $5.4 \:\mathrm{dB/cm}$ at $100 \:\mathrm{GHz}$ to $90 \:\mathrm{dB/cm}$ at $1 \:\mathrm{THz}$. The fairly low attenuation just above the minimum frequency for pair-breaking suggests that the $0.9\:\mathrm{cm}$ filters used in previous works may be too short to absorb all pair-breaking photons \cite{serniak_direct_2019, diamond_distinguishing_2022}. To test this, we measured the ground-state parity switching rate of our device during several different cooldowns while varying the length of Eccosorb filtering inside the final stage of device shielding (see table \ref{rate table}). We find that the parity switching rate decreases as we increase the length of the Eccosorb filter. The first $0.9\:\mathrm{cm}$ of Eccosorb is most effective -- compared to not having any filter in the final shield, it reduces the parity switching rate by $26\:\mathrm{dB}$. Adding longer filters resulted in continued improvement, but the attenuation of pair-breaking photons was reduced to roughly $5\:\mathrm{dB/cm}$ after the first $0.9\:\mathrm{cm}$ of filtering. We never observe the parity switching rate to saturate as we continue to add more Eccosorb, suggesting that the parity switching rate may still be dominated by photons when the qubit is in its the ground state, even with the longest Eccosorb filters that we measured\footnote{As explained in the main text, when the qubit is in the \textit{excited} state, the parity switching rate is vastly dominated by the resident QPs when using the optimal filtering and shielding configuration.}.

The observed filter length dependence of the ground-state parity switching rate can be explained by high temperature radiation incident on the filter (at least several $\mathrm{K}$).
The Eccosorb may be filtering blackbody radiation from the $4\:\mathrm{K}$, $50 \:\mathrm{K}$, or $300\:\mathrm{K}$ stage of the dilution refrigerator. Since the Eccosorb is more effective at higher frequencies, even a short filter is sufficient to eliminate most of the pair-breaking photons far above $100 \:\mathrm{GHz}$. Once the higher-frequency photons are eliminated, all that remains are the photons for which the attenuation length is the longest -- those around $100 \:\mathrm{GHz}$. The improvement we see at these long attenuator lengths is roughly consistent with the attenuation coefficient of Eccosorb near $100\:\mathrm{GHz}$.
 
All measured configurations have $0.9\:\mathrm{cm}$ long Eccosorb filters attached to the mixing chamber plate on the input line, output line, and SPA pump line (see Fig.~\ref{fig:wiring}). These filters reside outside our final shield, so although they may block photons from higher stages of the refrigerator, stray photons inside our still shield may leak into the lines after these filters. To protect the seams of our final layer of Eccosorb filtering, we enclose the final filters and the cavity itself inside a light-tight final shield.

\subsubsection{Device shielding}

The dilution refrigerator used in this experiment has built-in light shielding at the $300\:\mathrm{K}$, $50\:\mathrm{K}$,  $4\:\mathrm{K}$, and still stages. In addition to this built-in shielding, we enclose the aluminum cavity package inside a final custom-made shield mounted to the mixing chamber plate. All microwave lines enter this shield through a connectorized flange, and all seams in this shield are sealed with indium O-rings, except in the cooldowns noted in Table \ref{rate table}. The interior walls of the shield are painted with an absorptive coating of 86\% Stycast 2850 FT, 7\% catalyst 23LV, and 7\% carbon powder. We tested several different variants of this shield, along with different lengths of Eccosorb filter, as described in Table \ref{rate table}. Although there is significant cooldown-to-cooldown variation in the parity switching rate, we see the overall trend that once the shield is light-tight, adding additional Eccosorb to the final shield continually improves the parity lifetime. 

The results from improving the light-tightness of the shield are less conclusive than the improvement we get from adding more Eccosorb, but compared to the open-top shield, which appears to saturate as more Eccosorb is added, our indium-sealed shield continues to see improvement as we add more Eccosorb filters. We hypothesise that in the open-top shield, the parity switching rate is dominated by photons which leak into the shield after the final stage of Eccosorb filtering, resulting in no improvement from increasing the length of that filter. 

\begin{table}[h]
\centering
\begin{tabular}{|M{3cm} || M{1cm} | M{1cm} | M{1cm} | M{1cm} | M{1cm} |}
    \hline
    \backslashbox{shield}{filter}
     & $0  \:\mathrm{mm}$ 
     & $9  \:\mathrm{mm}$ 
     & $26 \:\mathrm{mm}$ 
     & $36 \:\mathrm{mm}$ 
     & $63 \:\mathrm{mm}$\\
    \hline
    open-top shield & 10,000$^\star$ & 360$^\star$ & 560$^\star$ & & \\
    \hline
    \multirow{2}{4em}{aluminum tape seal}   & 25,000 & 59 & 22 & & \\
                                            & & 53$^\bigtriangledown$ &  &  &  \\
    \hline
    \multirow{3}{4em}{indium O-ring seal} & & & 9.1 & 1.9 & 1.3$^{\times}$\\
                                          & & & &0.7$^{\square}$ & 1.2$^{\times }$\\
                                          & & & & & 0.14\\
    \hline 
\end{tabular}
\caption{
Parity switching rate for transmon in the ground state, in $\mathrm{sec}^{-1}$, for different filtering and shielding configurations. There is a clear trend of decreasing parity switching rate as we increase the length of the Eccosorb filters. Compared to no Eccosorb filters inside the final shield, including $63\:\mathrm{mm}$ of Eccosorb in this shield reduces the parity switching rate by a factor of well over $10^4$. The dependence of parity switching rate on the light-tightness of our shield is less clear, although we do see a modest improvement after fully sealing the shield with indium O-rings. The $36\:\mathrm{mm}$ filter was constructed by putting a $26\:\mathrm{mm}$ filter and a $10\:\mathrm{mm}$ in series. The $63\:\mathrm{mm}$ filter was constructed by putting $26\:\mathrm{mm}$ filter, a $10\:\mathrm{mm}$ filter, and three $9\:\mathrm{mm}$ filters in series. All filters were homemade, except for the $10\:\mathrm{mm}$ filter, which was manufactured by BlueFors.\\
$\star$ - data from Refs. \cite{serniak_direct_2019} and \cite{serniak_hot_2018}. It was taken in a different dilution refrigerator than the one used in the present experiment and in a radiation shield that was not light-tight.\\
$\bigtriangledown$ - covered SMA connectors inside the final shield with aluminum tape\\
$\square$ - added a piece of Eccosorb LS absorbing foam inside the final shield\\
$\times$ - packed the final shield completely with Eccosorb LS absorbing foam. We later discovered that the Eccosorb foam was interfering mechanically with the sealing of the indium O-ring. {This might have compromised the light-tightness of the shield during these cooldowns}}
\label{rate table}
\end{table}

\begin{figure}[!htbp]
\label{tau vs ecco plot}
\includegraphics[scale=1.0]{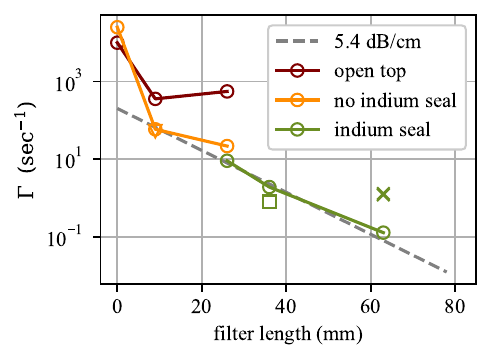}
\caption{Plot of the data shown in Table~\ref{rate table}. Lines are guides for the eye. Unfilled circles represent data points with no superscript in Table~\ref{rate table}, other symbols correspond to the superscript shown in Table \ref{rate table}. The gray dashed line corresponds to the minimum measured attenuation of Eccosorb CR-110 with an arbitrary vertical offset. We expect the parity switching rate to scale at this rate for a sufficiently long filter. 
}
\end{figure}

\subsection{System parameters}
The frequency of our readout resonator is $f_r = 9.201\:\mathrm{GHz}$. The transmon frequency is $f_q = 3.826\:\mathrm{GHz}$. The Josephson and charging energies of the transmon are $E_J/h = 6.24\:\mathrm{GHz}$ and $E_C/h = 357\:\mathrm{MHz}$, respectively. The relaxation time of the transmon is $T_1 = 193 \pm 25\:\mathrm{\mu s}$ and the dephasing time is $T_2^\star = 5
 \pm 1
\:\mathrm{\mu s}$ at the offset charge $n_g = 0$.

\subsection{Transmon readout}
\label{sec:readout}
Here, we describe the procedure that we use for reading out the state of our transmon qubit. In a standard fashion, the transmon is embedded into a 3D aluminum cavity readout resonator \cite{paik_observation_2011}. The capacitive coupling between the transmon and the readout resonator shifts the resonator frequency by an amount which depends on the transmon state \cite{koch_charge-insensitive_2007}. The latter can thus be measured by monitoring the signal reflected from the resonator. An unusual feature of our system is that the resonator dispersive shifts are appreciably different not only for the ground and the excited state of the transmon \cite{koch_charge-insensitive_2007} but also for its two possible charge-parities \cite{serniak_direct_2019}. This allows us to directly readout both the plasmon and the parity state of the transmon in a single shot.

The dependence of dispersive shifts on parity deserves an explanation. Indeed, this dependence does not reduce to the quantum capacitance  $\propto \partial^2_{n_g} E$ of the transmon, since the charge dispersion of its ground and excited state energies is weak.
Rather, it comes from the strong hybridization between the readout resonator and $|0\rangle \leftrightarrow |3\rangle$ and $|1\rangle\leftrightarrow|4\rangle$ transitions of the transmon \cite{serniak_direct_2019}. The hybridization is strong because the frequency of these transitions is close to the resonator frequency. Pronounced dependence of the transition frequencies on the offset charge $n_g$ leads to the parity-dependence of the hybridization and thus to that of the dispersive shifts.

\begin{figure}[h]
\includegraphics[scale=1.0]{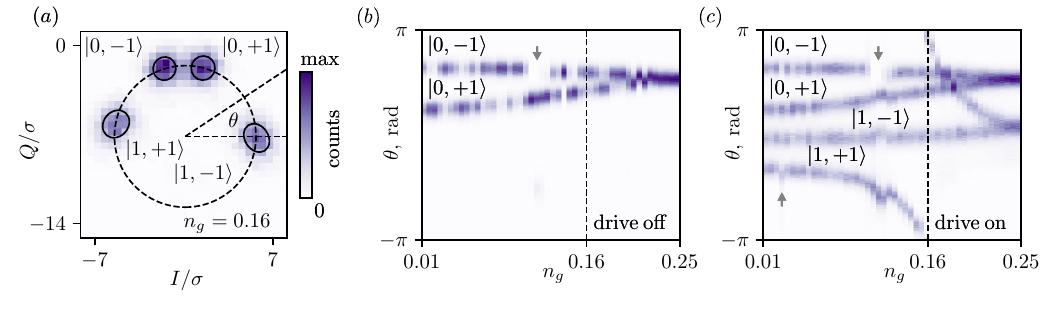}
\caption{
Readout histograms at different values of the offset charge $n_g$. Individual measurements are obtained by sending a readout pulse to the sample and measuring the reflected signal. The carrier frequency of the readout pulses corresponds to the (dressed) resonator frequency $f_r = 9.201\:\mathrm{GHz}$; each pulse is $11\:\mathrm{\mu s}$ long. 
(a) Readout histogram at a particular value of the offset charge, $n_g = 0.16 \pm 0.003$, {close to the one used in the main text ($n_g = 0.163$)}. The measurements are obtained as a sequence of $30$-seconds long time traces in which the readout pulse is played every $2\:\mathrm{ms}$. Every $11\:\mathrm{\mu s}$ between the measurements a parity-unselective $\pi$-pulse is played on the qubit to scramble its state between the ground and the excited state. Four distributions in the histogram correspond to different transmon states $|i,\mathcal{P}\rangle$ where $i$ denotes the plasmon state and $\mathcal{P}$ denotes the charge-parity. (b) The offset-charge dependence of the readout angle $\theta$ [see panel (a)] in the absence of the qubit-scrambling pulses. The transmon resides in the ground state in one of the two parity sectors. Arrow close to $n_g = 0.12$ points to a feature in the readout histogram caused by the readout-induced heating. Dashed line corresponds to the value of $n_g=0.16$ used in panel (a). (c) The offset-charge dependence of the readout in the presence of the scrambling pulses. Second readout-induced feature is visible in state $|1,+1\rangle$, see arrow around $n_g = 0.03$.
\label{fig:S1}}
\end{figure}

We demonstrate the joint readout of the plasmon and the parity state of the transmon in Fig.~\ref{fig:S1}. The figure shows the readout histograms for different values of the offset charge $n_g$ with and without a pulse scrambling the qubit state \cite{serniak_direct_2019}. The offset charge is measured spectroscopically \cite{serniak_hot_2018} and controlled with voltage applied to the resonator coupling pin (for details of $n_g$ control see Section~\ref{sec:msmt-prot}). Without the scrambling pulses, the qubit spends the vast majority of its time in the ground state with either value of the charge-parity, $\mathcal{P}=\pm 1$ [see Fig.~\ref{fig:S1}(b)]. Upon the application of the scrambling pulses, the first excited state of the transmon acquires an appreciable population for both parities [see Fig.~\ref{fig:S1}(c)]. The latter experiment shows that for a broad range of $n_g$ all four possible states are discernible \cite{serniak_direct_2019}. Notice that at $n_g = 0.03$ and $n_g = 0.12$ (in units of $2e$) the readout histograms are distorted [purple arrows in Fig.~\ref{fig:S1}(c)]. These distortions appear because at these values of $n_g$ our high-power readout pulses excite the qubit to high-energy states \cite{sank_measurement-induced_2016}. The large power of the readout pulses is required to compensate for the insertion loss of the Eccosorb filter (about $13\:\mathrm{dB}$ at room temperature).

For the experiments presented in the main text we picked a particular value of the offset charge $n_g = 0.163 \pm 0.003$ at which all four relevant states can be resolved. This choice was based on a trade-off between the distinguishability of the different states and the amount of qubit heating by the readout pulses. As described in the main text, at the chosen value of $n_g$ the measured population of the first excited transmon state does not exceed $p_\mathrm{exc} = 0.2\%$ [which corresponds to the effective qubit temperature of $T_q = 29\:\mathrm{mK}$]. At values of $n_g$ with a better readout signal-to-noise ratio, the measured excited state population is higher. For example, at $n_g = 0.06$ the excited state population is $p_\mathrm{ext}\sim 1\%$. We believe that the variation of the excited state population with the offset charge arises due to the proximity of the readout tone to various multi-photon resonances in the transmon spectrum. As discussed in Section~\ref{sec:ng-indep}, the choice of the offset charge is not consequential for the extracted parity-switching rates.
{\subsection{Qubit fabrication}
Our transmon qubit consists of two patterned aluminum films with thicknesses $20\:\mathrm{nm}$ and $30\:\mathrm{nm}$. The films are deposited on a c-plane sapphire substrate in a single double-angle deposition. The patterning was done using PMMA/MMA bilayer resist via the standard electron-beam lithography. The Josephson junction was formed using the bridge-free technique. More details of the qubit fabrication can be found in Ref.~\cite{serniak_direct_2019} which used the same device.}

\section{Measurement and data analysis}
\label{sec:msmt-data-an}
As described in the main text, the final result of our measurement are the parity-switching rates in the ground and the excited state of the transmon qubit, $\Gamma_0$ and $\Gamma_1$, respectively. The two rates are obtained by first measuring the parity-switching rate $\Gamma$ as a function of the controlled excited state population of the qubit, $p_1$, and then extrapolating the rate via $\Gamma = (1 - p_1) \Gamma_0 + p_1 \Gamma_1$. Here, we provide the details of how $p_1$ and $\Gamma$ are measured, see Sections~\ref{sec:epop} and \ref{sec:rates}, respectively.

\subsection{Measurement protocol}
\label{sec:msmt-prot}
We begin by broadly outlining the measurement protocol. In our measurement, we collect a sequence of 30-second long qubit readout jump traces. Each such trace is obtained by reading out the state of the transmon once every $2\:\mathrm{ms}$ with a $11\:\mathrm{\mu s}$-long readout pulse (see Section~\ref{sec:readout} for the details of the readout). Transmon measurements are interleaved with short ($\sigma = 8\:\mathrm{ns}$) parity-unselecive Gaussian pulses of a variable amplitude played every $11\:\mathrm{\mu s}$. The role of these pulses is to produce the controllable steady-state population of the excited qubit state $p_1$ similar for both parity sectors (see Section~\ref{sec:epop} for the calibration).

All of our measurements are performed at a fixed value of the offset charge, $n_g = 0.163\pm0.003$. To ensure the consistency of the offset charge, before collecting each measurement trace, we measure $n_g$ through Ramsay spectrometry \cite{riste_millisecond_2013} and correct it by applying the gate voltage to the resonator coupling pin (see Fig.~\ref{fig:wiring}). After collecting the trace, we measure the offset charge again. If $n_g$ deviates from $0.163$ by more than $0.003$ due to the charge drift, we discard the trace.

\subsection{Extraction of the excited state probability}
\label{sec:epop}
We analyze the excited state population $p_1$ by fitting the distribution of the measurement outcomes to a Gaussian mixture model, see Fig.~\ref{fig:S2}(a). The model approximates the measured distribution with a weighted combination of several Gaussian distributions. The fitting is performed using the Pomegranate package in Python \cite{schreiber_pomegranate_2018}.

For the fitting, we choose a model with five distributions. Four distributions correspond to the transmon states, $|0,\mathcal{P}\rangle$ and $|1,\mathcal{P}\rangle$ with $\mathcal{P}=\pm 1$, and the fifth distribution accounts for the nonzero population of higher excited states of the transmon {(see grey ellipse in Fig.~\ref{fig:S2})}. To compute the total excitation probability $p_1$, we add up the probabilities of the three excited states in the model.
\subsection{Extraction of the parity switching rates}
\label{sec:rates}
To extract the parity switching rate for a given temperature and qubit drive strength, we fit the measured jump traces to a hidden Markov model \cite{schreiber_pomegranate_2018}. Here, we outline the details of our fitting procedure.

To begin with, we note that in our case fitting to a Hidden Markov model is complicated by several factors: (i) our data is two-dimensional; (ii) it involves switching between four transmon states; (iii) switching occurs at various timescales, i.e., switching between the ground and the excited state induced by the drive is much quicker than the parity switching. This makes fitting the raw data to a hidden Markov model difficult because of the large number of free parameters. To reduce the number of free parameters, we pre-process the data to simplify the hidden Markov analysis and make it more robust.

\begin{figure}[h]
\includegraphics[scale=1.0]{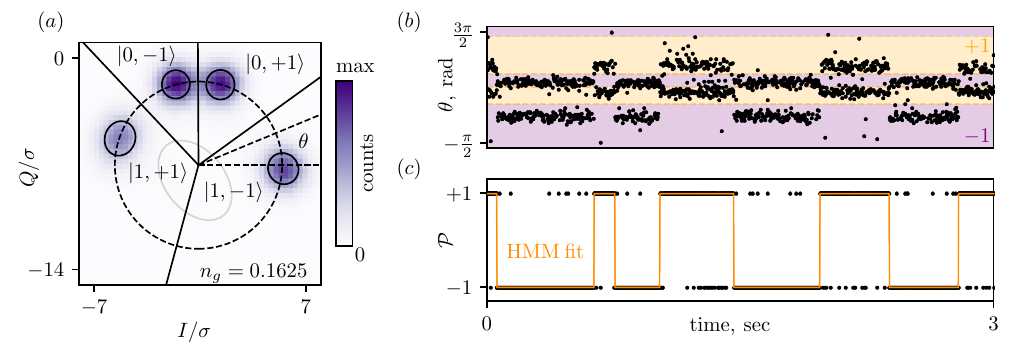}
\caption{
Measurement of the dependence of the parity switching rate on the excited state population. The transmon is read out once every $2\:\mathrm{ms}$. The measurements are interleaved with variable-amplitude qubit excitation pulses played every $11\:\mathrm{\mu s}$. (a) Histogram of measurement outcomes. The excited state population is computed by fitting the histogram to a Gaussian mixture model. The distributions corresponding to different qubit states within this model are shown as black ellipses; grey ellipse is an extra distribution which accounts for short jumps to highly-excited transmon states. Angle $\theta$ parametrizes measurement outcomes and is used for state assignment. The latter is done according to thresholds shown with solid lines. (b) Jump trace of measured $\theta$. Rare abrupt jumps in the trace correspond to the parity switching. Quick jumps within each parity correspond to switching between the ground and the excited state induced by the applied drive. Orange and purple regions show which measurements are assigned to parity $+1$ and which to parity $-1$. (c) The parity trace based on the state assignment applied to the measurement trace in (b). The vast majority of flips of the assigned parity occur due to the measurement errors. To obtain a reliable information about the parity switching we fit the parity trace to a hidden Markov model (solid orange line).
\label{fig:S2}}
\end{figure}

The main idea of the pre-processing is to map the noisy two-dimensional continuous-variable jump trace to a noisy jump trace of a \textit{single discrete} variable corresponding to the charge-parity. This procedure proceeds as follows. First, we analyze the histogram of measurement outcomes in the $I-Q$ plane, see Fig.~\ref{fig:S2}(a). The four distributions in the figure correspond to different states of the transmon, $|0,\mathcal{P}\rangle$ and $|1,\mathcal{P}\rangle$ with parity $\mathcal{P}=\pm1$. As usual, the means of the distributions fall on a circle. We parameterize the circle with an angle $\theta$. A representative jump trace of $\theta$ is shown in Fig.~\ref{fig:S2}(b). Next, we divide the circle into sectors corresponding to four transmon states. The boundaries of the sectors are formed by the bisectors between the adjacent distributions. Depending on the sector in which $\theta$ falls for a given measurement, the measurement can be assigned to either $\mathcal{P} = +1$ or $\mathcal{P}=-1$. The jump trace of $\mathcal{P}$ obtained in this way is shown in Fig.~\ref{fig:S2}(c).

Next, we fit the thresholded parity measurements to a two-state discrete-variable hidden Markov model [see dashed lines in Fig.~\ref{fig:S2}(c)]. As a result, we obtain two rates: rate of parity switching from $+1$ to $-1$, $\Gamma_{+ \rightarrow -}$, and rate of parity switching from $-1$ to $+1$, $\Gamma_{- \rightarrow +}$. We also obtain the error rates for the parity measurements both for $\mathcal{P}=+1$ and $\mathcal{P}=-1$. These four parameters completely determine the hidden Markov Model. We then take the average $\Gamma = (\Gamma_{+\rightarrow -} + \Gamma_{-\rightarrow +})/2$, which we call the parity switching rate. For measurements presented in Figs.~2~and~3 of the main text, we always find that $\Gamma_{+ \rightarrow -}$ is close to $\Gamma_{- \rightarrow +}$. We note that the parity measurement error rate is much higher when the system is in one parity state compared to the other, as seen in Fig.~\ref{fig:S2}(b). This is likely due to readout-induced heating events. This heating is taken into account in the hidden Markov model as a measurement error. It does not affect the extracted rates since the measurement repetition rate is much faster than the parity switching rate.

Repeated application of the described protocol for different qubit drive strengths allows us to map out the dependence of the parity switching rate $\Gamma$ on the excited state population $p_1$ and thus extract the parity switching rate in the ground and the excited state of the transmon, $\Gamma_0$ and $\Gamma_1$. This way we produce Figure 3 of the main text of the manuscript.

\subsection{Details of the parity manipulation experiment}
{Here, we provide the details of the parity manipulation experiment, Figure~4 of the main text. In short, to control the parity, we apply a drive at frequency $f_q^+$ (see below for the details of the pulse sequence). As a proxy for the drive strength, we use the steady-state probability $p_e[+1]$ of the excited qubit state conditioned on parity being $+1$. For weak driving, $p_e[+1]$ is proportional to the drive power due to qubit relaxation; for strong driving, $p_e[+1]$ saturates to $1/2$ due to Rabi oscillations. Then we measure the probability of the ground state with parity $-1$ as a function of $p_e[+1]$. Measurement results are shown as teal points in Fig.~4 of the main text. Upon the application of the drive, the parity becomes biased towards $-1$. The degree of parity polarization quickly increases with $p_e[+1]$. The direction of polarization can be reversed by driving at $f_q^-$ instead of $f_q^+$ (see orange points in Fig.~4).}

\subsubsection{Theoretical model}
{To describe the data of Fig.~4 theoretically, we apply detailed balance considerations. When we drive the system at $f_q^+$, the total rate of going from parity $+1$ to the ground state with parity $-1$ is $(1 - p_e[+1])\Gamma_0 + p_e[+1] \Gamma_1$. The rate of going from the ground state with parity $-1$ to $+1$ is $\Gamma_0$, since the drive does not excite the qubit if parity is $-1$. Then, from a rate equation, we find the steady-state probability of the ground state with parity $-1$,
\begin{equation}
   \label{eq:4}
    p[- 1] = \frac{1}{2}\left(1 - \frac{(\Gamma_1/\Gamma_0-1)p_e[+1]}{2+(\Gamma_1/\Gamma_0-1)p_e[+1]}\right).
\end{equation}
An analogous equation can be obtained for driving at $f_q^{-}$. Using $\Gamma_1 / \Gamma_0$ as a fitting parameter, we extract $\Gamma_1 / \Gamma_0 =37 \pm 2$ consistent with the data of Fig.~3 of the main text.}

\subsubsection{Detailed protocol of the measurement}
{Practically, instead of using continuous driving, we excite the qubit by using frequent selective excitation Gaussian pulses with $\sigma = 400\:\mathrm{ns}$. For such long pulses, the linewidth becomes small compared to charge dispersion $\delta f_q \sim 10\:\mathrm{MHz}$. This implies that if the pulse is resonant for one of the parities of the transmon, it is strongly off-resonant for another parity. The measurement proceeds as follows. First, by applying a DC voltage to the resonator coupling pin, we set the offset charge to a particular convenient value, $n_g = 0.163$ (see Sections~\ref{sec:readout} and~\ref{sec:msmt-prot} for discussion). Then, every $13\:\mu s$ we apply a variable-amplitude selective pulse resonant with the qubit transition for one of the parity sectors. Every $2.4\:\mathrm{ms}$ we measure the qubit state. Finally, every 36 seconds we verify that $n_g$ did not drift. If since the last measurement $n_g$ drifted by more than $0.003$, we discard the past 36 seconds of measurements and correct $n_g$. Given a set of measurement outcomes, we compute the populations of the different transmon states by fitting the data to a Gaussian mixture model (akin to how it was done in Section~\ref{sec:epop}).}
\subsection{Nuances of the measurement}
\subsubsection{Independence of the rate on the choice of the offset charge}
\label{sec:ng-indep}
The parity switching rates in our experiments should not depend on the particular value of the offset charge picked for the measurement. This is because the charge dispersion of transmon energy levels ($|0\rangle$ and $|1\rangle$ in particular) is much smaller than the temperature. To make sure that this is actually the case, we measure the dependence of the parity switching rate on the excited qubit state population for three different values of $n_g$, see Figure~\ref{fig:consistency}(a). Clearly, the rates measured at the different values of the offset charge differ only within the measurement error bar. This justifies the use of a particular convenient value of $n_g$ in our experiment.
\subsubsection{Repetition rate}
In our work, we read out the transmon sparsely, only once every $\sim 2\:\mathrm{ms}$ (the readout tone is turned off for $99.5\%$ of the measurement time). The reason for this is twofold. First, reading out sparsely allows us to avoid transmon heating by the readout pulses. Reading out more frequently can systematically bias the extracted parity switching rates. Indeed, readout-induced heating can excite the transmon to one of its high-energy states, for example, $|3\rangle$ or $|4\rangle$. The rates in these excited states can be very different from $\Gamma_0$ and $\Gamma_1$. Therefore, the measured rates would acquire uncontrollable and potentially large offsets from their true values. Second, frequent readout can directly increase the QP tunneling rate via photon-assisted processes. In such processes, photons residing in the readout resonator can force the QPs to tunnel and give them energy to traverse the gap difference at the junction. In our experiment, this processes are especially efficient since the resonator frequency $f_r = 9.201\:\mathrm{GHz}$ is higher than the gap difference, $\delta\Delta = 4.52\:\mathrm{GHz}$. Based on theory estimates, we expect that reading out continuously at the power of our readout pulses would lead to a change in $\Gamma_0$ larger than the ``bare'' rate measured with a sparse readout.

\begin{figure}[h]
\includegraphics[scale=1.0]{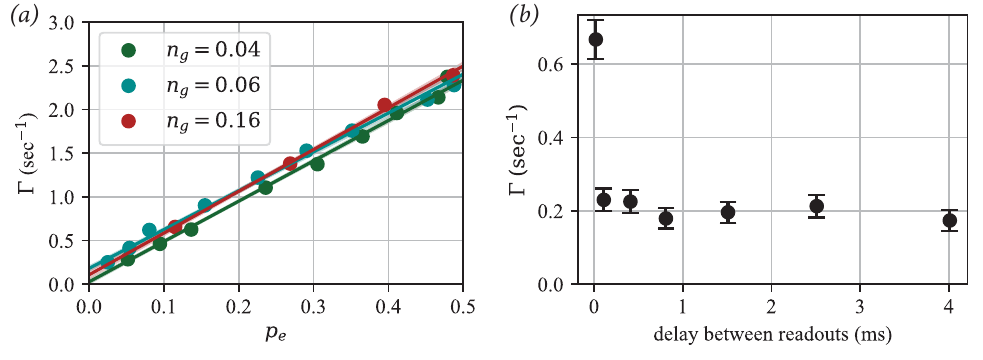}
\caption{
Sanity checks for the measurements presented in the main text. (a) The independence of the measured parity switching rates on the offset charge. The measurement here is conceptually similar to that presented in Figure 3(a) of the main text of the manuscript. Points with different colors correspond to different values of the offset charge. (b) The dependence of the extracted parity switching rate on the delay time between the qubit measurements. {The measurement is carried out at $n_g = 0.163$ (in units of $2e$).} Rate is extracted from the measured jump traces by fitting their power spectrum to a Lorentzian \cite{riste_millisecond_2013}. At small delay, the measurement is biased and gives an incorrect value of the rate (see discussion in the text). At the delay of $2\:\mathrm{ms}$ -- value of the delay used in the main text of the paper -- the bias is eliminated.
\label{fig:consistency}}
\end{figure}

To make sure that the delay between the readout pulses is long enough to avoid the bias, we measure how the extracted parity switching rate depends on the delay, see Figure~\ref{fig:consistency}(b). This plot shows that the delay of $2\:\mathrm{ms}$ that we use in the main text is clearly enough to suppress the bias.

We note that sparse readout can make the extraction of the parity switching rates unreliable if the delay between the measurements becomes comparable to the parity-switching time. In our experiments, however, the parity switching time is much larger than the measurement delay. We believe that this justifies our approach.
\subsubsection{Thermalization time}
In our measurement of the temperature dependence, Fig.~3 of the main text, we add a time delay between the measurement of subsequent points. This delay is needed for the system to thermalize to a new temperature. To estimate the required wait time we quickly switched the temperature of the mixing chamber plate from $80\:\mathrm{mK}$ to $20\:\mathrm{mK}$ and monitored how the measured parity switching rate -- which is indicative of the QP temperature -- changes with time. From this measurement we see that four hours is required to largely thermalize the system. We thus wait four hours after changing the temperature before starting a new measurement.

\section{Theoretical model and data fitting}
In this section, we outline the details of our theoretical model. In Subsections~\ref{sec:qpt} and \ref{sec:approx} we derive the expressions for the QP tunneling rates. In Section~\ref{sec:fit}, we describe the subtleties of the data fitting procedure that we use. In Subsection~\ref{sec:kin} we discuss subtle complications related to the QP kinetics and show that for our system they can be neglected. Finally, in Subsection~\ref{sec:pat} we discuss the photon-induced pair breaking. In particular, we derive the relation between the pair-breaking rates in the ground and in the excited qubit states.

\subsection{Quasiparticle tunneling rates}
\label{sec:qpt}
We start by computing the QP tunneling rates. These derivations closely follow previous works \cite{glazman_bogoliubov_2021, catelani_relaxation_2011, catelani_quasiparticle_2011, catelani_decoherence_2012, diamond_distinguishing_2022, marchegiani_quasiparticles_2022}. In the vicinity of the junction, we model the transmon as two superconductors coupled with a point-like tunnel junction. We take superconductivity into account within the framework of the Bardeen-Cooper-Schrieffer theory. We assume that superconducting gaps in the two superconductors are different, $\Delta$ for the left superconductor and $\Delta + \delta\Delta$ for the right superconductor. The Hamiltonian of the system within this model is given by 
\begin{gather}
H	=\sum_{k\sigma}\epsilon_{L,k}a_{L, k\sigma}^{\dagger}a_{L, k\sigma}+\sum_{k\sigma}\epsilon_{R,k}a_{R, k\sigma}^{\dagger}a_{R, k\sigma}+H_{T}+H_{C},\quad \epsilon_{L,k} = \sqrt{\Delta^2 + \xi_k^2},\quad \epsilon_{R,k} = \sqrt{[\Delta+\delta\Delta]^2 + \xi_k^2},\notag \\
H_{C}=4E_{C}\left(N - n_{g} + \frac{1}{4}P\right)^{2},\quad
H_{T}	=\sum_{kk^{\prime}\sigma}\frac{t}{V}(e^{\frac{i\varphi}{2}}\tilde{a}_{L, k\sigma}^{\dagger}\tilde{a}_{R, k^{\prime}\sigma}+\mathrm{h.c.}).\label{eq:ham}
\end{gather}
Here, $\sigma$ denotes spin, $k$ denotes the wave vector, and $\xi_k$ denotes the kinetic energy (relative to Fermi level). $a_{L,k\sigma}$ and $a_{R,k\sigma}$ are the fermionic operators of superconducting QPs; $\epsilon_{L,k}$ and $\epsilon_{R,k}$ are the corresponding energies. In the expression for $H_C$, $N$ is the number of tunneled Cooper pairs, $E_C$ is the charging energy, and $n_g$ is the offset charge (in units of $2e$). $P=\pm 1$ is the fermion parity of the transmon; tunneling of a Cooper pair preserves the parity while the tunneling of a QP flips it. Parameter $t$ characterizes the strength of tunneling between the superconductors. $V$ is the normalization volume; it drops from the final results. Finally, operators $\tilde{a}_{L, k\sigma}$ and $\tilde{a}_{R, k\sigma}$ correspond to electrons in the absence of superconductivity. They can be related to the QP operators as
\begin{equation}
\label{eq:bare-qp}
\left(\begin{array}{c}
\tilde{a}_{R, k\sigma}\\
\tilde{a}_{R, -k,-\sigma}^{\dagger}
\end{array}\right)	=\left(\begin{array}{cc}
u_{R,k} & -\sigma v_{R,k}\\
\sigma v_{R,k} & u_{R,k}
\end{array}\right)\left(\begin{array}{c}
{a}_{R,k\sigma}\\
{a}_{R,-k,-\sigma}^{\dagger}
\end{array}\right),\quad u_{R,k}^{2}=\frac{1}{2}\left(1+\frac{\xi_{k}}{\epsilon_{R,k}}\right),\quad v_{R,k}^{2}=\frac{1}{2}\left(1-\frac{\xi_{k}}{\epsilon_{R,k}}\right)
\end{equation}
and similarly for $\tilde{a}_{L, k\sigma}$. The Josephson energy of the transmon can be related to the tunneling constant $t$ as \cite{marchegiani_quasiparticles_2022}
\begin{equation}
\label{eq:ej2}
E_{J} = \pi^2\left(\Delta + \frac{\delta\Delta}{2}\right) t^{2}\nu_{0}^{2},\quad \delta\Delta\ll \Delta.
\end{equation}
{In this expression $\nu_0$ is the normal-state density of states at the Fermi level per spin projection.}
For the parameters of our experiment the further corrections to equation~\eqref{eq:ej2} due to the finite gap difference give a correction of only $\sim 0.1\%$ to $E_J$.
\subsubsection{Exact expressions for the quasiparticle tunneling rates}
Next, we compute the QP tunneling rates. The rate depends on the initial state of the qubit and can be broken down into contributions corresponding to different final qubit states. For the transmon qubit, initial state (denoted as $|i\rangle$) and final state ($|f\rangle$) are fully characterized by the parity $P$ and by the number of plasmon excitations. QP tunneling necessarily flips the parity, $P \rightarrow -P$. The plasmon state of the transmon, however, may or may not change depending on whether the QP exchanges energy with the qubit. QP tunneling events can be further characterized by whether the QP tunnels from the left superconductor to the right superconductor or vice versa. We distinguish these two types of events with a special index, $R\rightarrow L$ or $L\rightarrow R$. To summarize, we denote QP tunneling rates corresponding to different processes as either $\Gamma_{if}^{R\rightarrow L}$ or $\Gamma_{if}^{L \rightarrow R}$. To give an example, $\Gamma_{01}^{L\rightarrow R}$ denotes a transition in which a QP tunnels from the left superconductor to the right superconductor and the plasmon state changes from $|0\rangle$ to $|1\rangle$.

Next, we compute the rates using Fermi's Golden rule in which we treat the tunneling part of Hamiltonian \eqref{eq:ham}, $H_T$, as a perturbation. This allows us to generalize the results obtained in the limit of vanishing $\delta\Delta$ \cite{catelani_quasiparticle_2011, catelani_relaxation_2011, catelani_decoherence_2012} to the case with a finite $\delta\Delta$ \cite{diamond_distinguishing_2022, marchegiani_quasiparticles_2022}:
\begin{gather}
\label{eq:gammaLR}
\Gamma_{if}^{L\rightarrow R}	=16\frac{E_J}{h}\left(S_-^{L\rightarrow R}(f_{fi})|\langle f|\cos\frac{\varphi}{2}|i\rangle|^2 + S_+^{L\rightarrow R}(f_{fi}) |\langle f|\sin\frac{\varphi}{2}|i\rangle|^2\right),
\end{gather}
\begin{equation}
    \label{eq:sfLR}
    S_{\pm}^{L\rightarrow R}(f_{fi}) = \frac{1}{\Delta + \frac{\delta\Delta}{2}} \int_{\max(\Delta,\Delta +\delta\Delta + h f_{fi})}^\infty d\epsilon_\nu \frac{\epsilon_\nu(\epsilon_\nu-h f_{fi})\pm\Delta(\Delta+\delta\Delta)}{\sqrt{\epsilon_\nu^{2}-\Delta^{2}}\sqrt{(\epsilon_\nu-h f_{fi})^{2}-(\Delta+\delta\Delta)^{2}}}
\mathcal{F}_L(\epsilon_\nu)[1 - \mathcal{F}_R(\epsilon_\nu- h f_{fi})].
\end{equation}
In the equation for the structure factors $S_\pm$, Eq.~\eqref{eq:sfLR}, functions $\mathcal{F}_L(\epsilon_\nu)$ and $\mathcal{F}_R(\epsilon_\nu)$ are the distribution functions of the QPs in the left and in the right superconductor, respectively. These distribution functions encode the information about the number and the energy of the QPs.

For rates of QP tunneling from the right superconductor to the left superconductor we obtain expressions similar to Eq.~\eqref{eq:gammaLR} and Eq.~\eqref{eq:sfLR}: 
\begin{equation}
\label{eq:gammaRL}
\Gamma_{if}^{R\rightarrow L}	=16\frac{E_J}{h}\left(S_-^{R\rightarrow L}(f_{fi})|\langle f|\cos\frac{\varphi}{2}|i\rangle|^2 + S_+^{R\rightarrow L}(f_{fi}) |\langle f|\sin\frac{\varphi}{2}|i\rangle|^2\right),
\end{equation}
\begin{gather}
\label{eq:sfRL}
S_{\pm}^{R\rightarrow L}(f_{fi}) = \frac{1}{\Delta + \frac{\delta\Delta}{2}} \int_{\max(\Delta, \Delta + \delta\Delta-h f_{fi})}^\infty d\epsilon\frac{\epsilon_\nu(\epsilon_\nu+h f_{fi})\pm\Delta(\Delta + \delta\Delta) }{\sqrt{\epsilon_\nu^{2}-\Delta^{2}}\sqrt{(\epsilon_\nu+h f_{fi})^{2}-(\Delta+\delta\Delta)^{2}}}
\mathcal{F}_R(\epsilon_\nu+h f_{fi})[1 - \mathcal{F}_L(\epsilon_\nu)].
\end{gather}
The difference between the structure factors for different directions of tunneling is the sign near $h f_{fi}$ as well as swapped distribution functions.

The total rates of the QP tunneling in the ground and in the excited state of the transmon --- which we measure in our experiment --- can be expressed through the partial rates in Eqs.~\eqref{eq:gammaLR} and \eqref{eq:gammaRL}:
\begin{gather}
\label{eq:partial-rates}
    \Gamma_0^{\mathrm{qp}} = \Gamma_{00}^{R\rightarrow L} + \Gamma_{01}^{R\rightarrow L} + \Gamma_{00}^{L\rightarrow R} + \Gamma_{01}^{L\rightarrow R},\quad \Gamma_1^{\mathrm{qp}} = \Gamma_{10}^{R\rightarrow L} + \Gamma_{11}^{R\rightarrow L} + \Gamma_{12}^{R\rightarrow L} + \Gamma_{10}^{L\rightarrow R} + \Gamma_{11}^{L\rightarrow R} + \Gamma_{12}^{L\rightarrow R}.
\end{gather}
In Eq.~\eqref{eq:partial-rates} we neglected the transitions in which the transmon state changes by more than one excitation. This assumption is valid deep in the transmon limit ($E_J \gg E_C$) since the corresponding matrix elements of $\cos(\varphi/2)$ and $\sin(\varphi/2)$ are small.

\subsubsection{Distribution functions}
\begin{figure}[h]
\includegraphics[scale=1.0]{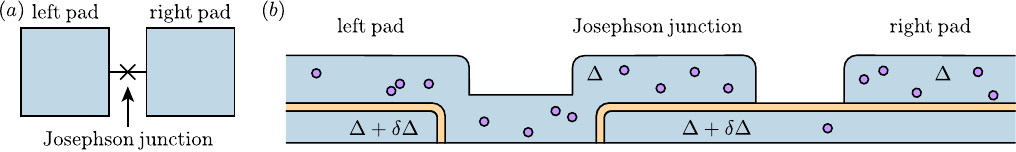}
\caption{
(a) Schematic of a transmon qubit. Two large superconducting pads are connected by a Josephson junction. (b) The junction is formed by overlapping two superconducting films through a layer of oxide. The two films have a different superconducting gap since they have a different thickness (20 nm for a high-gap film and 30 nm for a low-gap film). Transmon pads consist of both films. At small temperature, $k_B T \ll \delta\Delta$, QPs within each pad (purple) reside in the low-gap layer. At higher temperatures, the QPs start to spread into the high-gap layer.
\label{fig:junction-schematic}}
\end{figure}

In the calculation of the parity-switching rates, we use an expression for the QP distribution functions which is based on a set of assumptions. First, we assume that the distribution functions correspond to a non-equilibrium situation with a finite concentration of QPs even at small temperature. Second, we assume that the energy distribution of the QPs is well thermalized to the phonon bath despite their out-of-equilibrium density. This requires the time it takes for the QP to tunnel after its generation to be long compared to the timescale of inelastic electron-phonon processes \cite{kaplan_quasiparticle_1976, catelani_non-equilibrium_2019}. Since the transmon pads are large, the tunneling rate of individual QPs is small (of order of few seconds) making this assumption well-based. Third, we assume that QPs at the two sides of the junction are in equilibrium with each other, $\mathcal{F}_L = \mathcal{F}_R = \mathcal{F}$. The limits of validity of this assumption are discussed in Section~\ref{sec:kin}. Fourth, we assume that the rate of QP trapping (for example arising from vortices in the pads) exceeds the rate of QP recombination. Since the rate of recombination is proportional to the QP density, this assumption is valid at small enough QP densities [see Section~\ref{sec:kin} for discussion]. Finally, we note that both pads of our transmon qubit have regions with the low gap $\Delta$ and the high gap $\Delta + \delta\Delta$, see Figure~\ref{fig:junction-schematic}. We assume quick equilibration of QPs between the regions with different gaps within each pad. As a result, at small temperatures the QPs are mainly localized in the low-gap regions.

{To introduce the distribution function, let us denote the occupation probability of an eigenstate labeled with an index $\nu$ by $\mathcal{F}(\epsilon_\nu)$.} Under the listed assumptions, in both pads, $\mathcal{F}(\epsilon_\nu)$ can be expressed as
\begin{equation}
    \label{eq:distr}
    \mathcal{F}(\epsilon_\nu) = \zeta(T) \xqp^{\mathrm{res}} \sqrt{\frac{\Delta}{2\pi k_B T}}\exp\left(-\frac{\epsilon_\nu - \Delta}{k_B T}\right) + \exp\left(-\frac{\epsilon_\nu}{k_B T}\right).
\end{equation}
In Eq.~\eqref{eq:distr}, $T$ is the temperature of the phonon bath which we assume to be close to the mixing chamber temperature of our dilution refrigerator. $\xqp^{\mathrm{res}}$ is the \textit{total} number of resident QPs in a single transmon pad measured in units of total number of Cooper pairs in \textit{the low-gap film} of the pad. Dimensionless multiplier $\zeta(T)$ describes the delocalization of QPs from the low-gap film to the high-gap film within each pad upon the increase of temperature. When $\delta\Delta \ll \Delta$ -- which is the limit relevant for our data -- this multiplier reads\footnote{Expression \eqref{eq:eta} also requires the QP trapping rates in the two films within each transmon pad to be sufficiently close to each other, $|s_{\Delta+\delta\Delta} - s_\Delta| \exp\left(-\frac{\delta\Delta}{k_B T}\right)\ll s_{\Delta} $ (where $s_{\Delta+\delta\Delta}$ is the trapping rate in the high-gap film and $s_\mathrm{\Delta}$ is the trapping rate in the low-gap film).  We assume that this condition is fulfilled.}
\begin{equation}
    \label{eq:eta}
    \zeta(T) = \frac{V_\Delta}{V_\Delta+V_{\Delta + \delta\Delta}\exp\left(-\frac{\delta\Delta}{k_B T}\right)}
\end{equation}
where $V_{\Delta+\delta\Delta}$ and $V_{\Delta}$ are the volumes of the high-gap film and low-gap film in a single pad, respectively. At small temperatures, $k_B T \ll \delta\Delta$, $\zeta = 1$ implying that all QPs are localized in the low-gap region of the device {(we assumed $V_{\Delta+\delta\Delta} \sim V_\Delta$).}

\subsubsection{Evaluation of the structure factors}
Assuming the distribution function given by Eq.~\eqref{eq:distr} the structure factors in Eqs.~\eqref{eq:sfLR} and \eqref{eq:sfRL} can be calculated explicitly in the limit $k_B T \ll \Delta$ and $\delta\Delta \ll \Delta$.
Assuming that $\delta\Delta > |h f_{fi}|$ we find
\begin{gather}
    \label{eq:K0LR}
    S_+^{L\rightarrow R}(h f_{fi}) = \left(\zeta(T) \xqp^{\mathrm{res}}  
    \sqrt{\frac{\Delta}{2\pi k_B T}} + e^{-\frac{\Delta}{k_B T}} \right) e^{-\frac{\delta\Delta + h f_{fi}}{2k_B T}} K_0\left(\frac{\delta\Delta + h f_{fi}}{2k_B T}\right),\\
    \label{eq:K0RL}
    S_+^{R\rightarrow L}(h f_{fi}) =
    \left(\zeta(T) \xqp^{\mathrm{res}}  
    \sqrt{\frac{\Delta}{2\pi k_B T}} + e^{-\frac{\Delta}{k_B T}}\right)
    e^{-\frac{\delta\Delta + h f_{fi}}{2k_B T}} K_0\left(\frac{\delta\Delta - h f_{fi}}{2k_B T}\right),\\
    \label{eq:K1LR}
    S_-^{L\rightarrow R}(h f_{fi}) = \left(\zeta(T) \xqp^{\mathrm{res}}  
    \sqrt{\frac{\Delta}{2\pi k_B T}} + e^{-\frac{\Delta}{k_B T}}\right)
    \frac{1}{2} \frac{\delta\Delta + h f_{fi}}{\Delta} e^{-\frac{\delta\Delta + h f_{fi}}{2k_B T}} K_1\left(\frac{\delta\Delta + h f_{fi}}{2k_B T}\right),\\
    \label{eq:K1RL}
    S_-^{R\rightarrow L}(h f_{fi}) = \left(\zeta(T) \xqp^{\mathrm{res}}  
    \sqrt{\frac{\Delta}{2\pi k_B T}} + e^{-\frac{\Delta}{k_B T}}\right)
    \frac{1}{2} \frac{\delta\Delta - h f_{fi}}{\Delta} e^{-\frac{\delta\Delta + h f_{fi}}{2k_B T}} K_1\left(\frac{\delta\Delta - h f_{fi}}{2 k_B T}\right),
\end{gather}
where $K_0$ and $K_1$ are modified Bessel functions of the second kind. Similar expressions can be obtained for the opposite limit, $\delta\Delta < |h f_{fi}|$.

\subsection{Approximate expressions for the quasiparticle tunneling rates}
\label{sec:approx}

Assuming that the temperature of the system is small, $k_B T\ll \delta\Delta - h f_q$, it is possible to obtain simple equations for the total parity switching rates in the ground and excited transmon states, Eqs.~(2) and (3) of the main text. This is allowed by the fact that in the considered limit the structure factors in Eqs.~\eqref{eq:K0LR} -- \eqref{eq:K1RL} can be approximated by exponential functions.

We start by noting that as long as $E_J \gg E_C$ the cosine matrix element in Eq.~\eqref{eq:gammaLR} and Eq.~\eqref{eq:gammaRL} is only appreciable for the transitions that do not change the plasmon state of the qubit. In contrast, the sine matrix element is only appreciable if the plasmon state changes by one photon. In this case, we can express the matrix elements as
\begin{equation}
    \label{eq:me-approx}
    |\langle j|\cos\frac{\varphi}{2}|i\rangle|^2 = \delta_{ij} ,\quad |\langle j|\sin\frac{\varphi}{2}|i\rangle|^2 = \frac{1}{4}\sqrt{\frac{2E_C}{E_J}}(i\delta_{i-1,j} + (i+1)\delta_{i+1, j}).
\end{equation}
Calculating the structure factors with the outlined set of approximations and using Eq.~\eqref{eq:me-approx} we obtain the partial rates:
\begin{gather}
\Gamma_{00}^{R\rightarrow L} = \Gamma_{00}^{L\rightarrow R} = \Gamma_{11}^{R\rightarrow L} = \Gamma_{11}^{L\rightarrow R} = \frac{16 E_J}{h} \sqrt{\frac{\delta\Delta}{8\Delta}}
\xqp^\mathrm{res} e^{-\frac{\delta\Delta}{k_B T}}
\label{eq:gammas1}
\\
\Gamma_{01}^{R\rightarrow L}= \frac{4 E_J}{h} \sqrt{\frac{2E_C}{E_J}}\frac{1}{\sqrt{\delta\Delta-h f_{10}}}
\sqrt{\frac{\Delta}{2}}\xqp^\mathrm{res} e^{-\frac{\delta\Delta}{k_B T}},\quad \Gamma_{01}^{L\rightarrow R} = \frac{4 E_J}{h} \sqrt{\frac{2E_C}{E_J}} \frac{1}{\sqrt{\delta\Delta + h f_{10}}} \sqrt{\frac{\Delta}{2}}\xqp^\mathrm{res} e^{-\frac{\delta\Delta + h f_{10}}{k_B T}}
\label{eq:gammas2}
\\
\Gamma_{10}^{R\rightarrow L}= \frac{4 E_J}{h} \sqrt{\frac{2E_C}{E_J}}\frac{1}{\sqrt{\delta\Delta+h f_{10}}}
\sqrt{\frac{\Delta}{2}}\xqp^\mathrm{res} e^{-\frac{\delta\Delta}{k_B T}},\quad \Gamma_{10}^{L\rightarrow R} = \frac{4 E_J}{h} \sqrt{\frac{2E_C}{E_J}} \frac{1}{\sqrt{\delta\Delta - h f_{10}}} \sqrt{\frac{\Delta}{2}}\xqp^\mathrm{res} e^{-\frac{\delta\Delta - h f_{10}}{k_B T}}
\label{eq:gammas3}
\\
\Gamma_{12}^{R\rightarrow L} =\frac{8 E_J}{h} \sqrt{\frac{2E_C}{E_J}}\frac{1}{\sqrt{\delta\Delta-h f_{21}}}
\sqrt{\frac{\Delta}{2}}\xqp^\mathrm{res} e^{-\frac{\delta\Delta}{k_B T}},\quad \Gamma_{12}^{L\rightarrow R} = \frac{8 E_J}{h} \sqrt{\frac{2E_C}{E_J}} \frac{1}{\sqrt{\delta\Delta + h f_{21}}} \sqrt{\frac{\Delta}{2}}\xqp^\mathrm{res} e^{-\frac{\delta\Delta + h f_{21}}{k_B T}}
\label{eq:gammas4}
\end{gather}
Note that the detailed balance relation holds:
\begin{equation}
    \Gamma_{10}^{R \rightarrow L} + \Gamma_{10}^{L \rightarrow R} = e^{\frac{h f_{10}}{k_B T}}(\Gamma_{01}^{R \rightarrow L} + \Gamma_{01}^{L \rightarrow R}).
\end{equation}
As long as $\delta\Delta, h f_q \gg k_B T$, the rates in Eqs.~\eqref{eq:gammas1} -- \eqref{eq:gammas4} have a clear hierarchy:
\begin{equation}
    \Gamma_{10}^{L\rightarrow R} \gg \Gamma_{01}^{R\rightarrow L}, \Gamma_{10}^{R\rightarrow L}, \Gamma_{12}^{R\rightarrow L}, \Gamma_{00}^{R\rightarrow L}, \Gamma_{00}^{L\rightarrow R}, \Gamma_{11}^{R\rightarrow L}, \Gamma_{11}^{L\rightarrow R}\gg\Gamma_{01}^{L\rightarrow R}, \Gamma_{12}^{L\rightarrow R}.
\end{equation}
This allows us to approximate the parity switching rates in the ground and excited transmons states, Eq.~\eqref{eq:partial-rates}, as
\begin{equation}
\label{eq:01-approx}
    \Gamma_{0}^\mathrm{qp} = \Gamma_{00}^\mathrm{R\rightarrow L} + \Gamma_{00}^\mathrm{L\rightarrow R} + \Gamma_{01}^{R\rightarrow L},\quad \Gamma_{1}^\mathrm{qp} = \Gamma_{10}^{L\rightarrow R}.
\end{equation}
Substitution of Eqs.~\eqref{eq:gammas1} -- \eqref{eq:gammas3} into Eq.~\eqref{eq:01-approx} results in equations (2) and (3) of the main text.

\subsection{Data fitting procedure}
\label{sec:fit}
In this section, we describe the procedure used for fitting the temperature dependence of the parity-switching rates in the ground and in the first excited state of the transmon, see Fig.~3 of the main text. In short, we attribute the majority of the parity-switching events to the tunneling QPs. Therefore, we use Eqs.~\eqref{eq:gammaLR} - \eqref{eq:sfRL} to calculate the parity switching rates in the ground and in the excited state of the transmon, Eq.~\eqref{eq:partial-rates}. We compute the integrals in the structure factors, Eq.~\eqref{eq:sfLR} and Eq.~\eqref{eq:sfRL}, numerically.

This procedure, however, has several important nuances that have to be taken into account to correctly describe the data. First, in the data, the ground state parity switching rate saturates at small temperature. At the same time, $\Gamma_{0}^\mathrm{qp}$ given by Eq.~\eqref{eq:partial-rates} does not saturate and reduces to zero as the temperature is decreased (at zero temperature quasiparticles cannot overcome the gap difference at the junction). We conjecture that this discrepancy might arise due to the residual pair-breaking by stray high-frequency photons. Indeed, absorption of such photons might give a contribution to the parity switching rate \cite{houzet_photon-assisted_2019, diamond_distinguishing_2022} which does not depend on the temperature as long as photon energy is much higher than the thermal energy $k_B T$ of the device. Presence of such a contribution would indeed result in the saturation of the ground state rate at small temperature. This hypothesis is reinforced by the fact that the observed rate of parity switching in the ground state is roughly consistent with the expectation for photon-induced pair breaking based on the amount of filtering (see Sections~\ref{sec:filtering-configs} and \ref{sec:pat} for discussion). An alternative explaination for the saturation of $\Gamma_0$ is significantly elevated QP temperature. This explanation is unlikely since even after $\Gamma_0$ saturates $\Gamma_1$ continues to decrease with reducing $T$.

Practically, to take into account the saturation of the ground state rate at small temperature, we add a temperature independent offset to this rate, $\Gamma_0(T) = \Gamma_\mathrm{offset} + \Gamma_0^{\mathrm{qp}}(T)$, where $\Gamma_0^\mathrm{qp}(T)$ is given by Eq.~\eqref{eq:partial-rates}. We add a similar offset to the excited state rate\footnote{We note that the offset rate due to photon-induced pair-breaking is in general different for the ground and for the excited state, see Section~\ref{sec:pat} for discussion. For the amount of filtering that we have, we expect this difference to be inconsequential for describing our data because the parity switching in the excited state is limited by QP tunneling at all temperatures available in our measurement.}, $\Gamma_1(T) = \Gamma_\mathrm{offset} + \Gamma_1^{\mathrm{qp}}(T)$. However for this rate the presence of the offset is not consequential since at all relevant temperatures $\Gamma_1^{\mathrm{qp}}(T)\gg \Gamma_\mathrm{offset}$.

Next, we comment on an important nuance behind out fitting procedure which manifests at high temperatures. In the main text we assumed that the parity switching rate can be simply written as
\begin{equation}
\Gamma(T) = (1 - p_1) \Gamma_0(T) + p_1 \Gamma_1(T),
\end{equation}
where $p_1$ can be controlled by driving the qubit. This implicitly assumes that the qubit does not leave the subspace of states $|0\rangle$ and $|1\rangle$. This, however, is only an approximation and for $T \gtrsim 60\:\mathrm{mK}$ state $|2\rangle$ acquires an appreciable population.  The minimal extension of this expression to higher temperature takes into account the population of the second excited state:
\begin{equation}
\label{eq:incl-gamma2}
    \Gamma(T) = (1 - p_{exc}) \Gamma_0(T) + p_{exc} \left(\frac{\Gamma_1(T)}{1+e^{-\frac{h f_{21}}{k_B T}}} +  \frac{\Gamma_2(T) e^{-\frac{h f_{21}}{k_B T}}}{1 + e^{-\frac{h f_{21}}{k_B T}}}\right).
\end{equation}
Here, $p_{exc}$ characterizes the probability of exciting the qubit out of the ground state. Therefore, when we extrapolate the total parity switching rate $\Gamma$ to zero and unit $p_{exc}$, we in fact recover $\Gamma_0$ and the weighted combination of $\Gamma_1$ and $\Gamma_2$, respectively. These are the two rates plotted in Fig.~3 as $\Gamma_0$ and $\Gamma_1$. To compute $\Gamma_2$ we use
\begin{equation}
    \label{eq:}
    \Gamma_2 = \Gamma_\mathrm{offset}  + \Gamma_2^\mathrm{qp}, \quad \Gamma_2^\mathrm{qp} = \Gamma_{21}^{R\rightarrow L} + \Gamma_{22}^{R\rightarrow L} + \Gamma_{23}^{R\rightarrow L} + \Gamma_{21}^{L\rightarrow R} + \Gamma_{22}^{L\rightarrow R} + \Gamma_{23}^{L\rightarrow R},
\end{equation}
where we again neglect transitions in which the transmon state changes by two or more excitations.

We note that even for $p_{exc} = 1$ the weight of $\Gamma_2$ in the total rate is no more than $\sim 18\%$ (at $110\:\mathrm{mK}$). Given that $\Gamma_2$ is close to $\Gamma_1$ for the parameters of our fits, the difference between the actual $\Gamma_1$ and the result of our extrapolation is at most several percent. We believe that this justifies us in ignoring these complications in the main text of the manuscript.

\subsection{Kinetics of quasiparticles}
\label{sec:kin}
One of the main assumptions of our theory is that in the absence of qubit driving, the system resides in a quasi-equilibrium state. This means that although the QPs have an excess density, their energy distribution is well-thermalized to the phonon bath. {As long as the qubit mode is thermalized to the same bath, the QPs at the two sides of the junction can be assumed to be at equilibrium with each other.}

When the qubit is driven, the QPs at the two sides of the junction do not have to be in equilibrium with each other. Indeed, due to the coupling between the qubit and the QPs, the state of the latter in principle also deviates from its quasi-equilibrium. To give an example, assume that we constantly project the qubit to the excited state. There, the qubit can donate its energy to assist QP tunneling from the low-gap side of the junction to the high-gap side [see Fig.~1(a) of the main text]. Thus, if the generation-relaxation dynamics of the QPs is not fast enough, most of the QPs will eventually be transferred to one of the transmon pads. There would thus be a strong imbalance between the QP densities on the two sides of the junction.

As a consequence of such kinetic effects, a simple linear relation for the parity switching rate,  $\Gamma = (1 - p_1)\Gamma_0 + p_1 \Gamma_1$, might no longer hold (see example below). At the same time, in our data we do not see deviations from the linear dependence [see Fig.~3(a) of the main text]. This hints that in our system the generation-relaxation dynamics of the QPs is quick enough such that the QPs across the junction are in equilibrium with each other. In this section, we explore the QP dynamics in the presence of qubit driving and show explicitly when the two QP subsystems can be assumed to be in equilibrium.

{Throughout this section we assume that the temperature is small, $k_B T\ll h f_{21}, \delta\Delta$.} These assumptions allow us to make several important simplifications. First, we can neglect the presence of thermal QPs [second term in Eq.~\eqref{eq:distr}] and only consider the excess resident QPs. Next, we can neglect the excitation of the transmon to its second excited state. Finally, we can also assume that within each pad of the transmon the QPs always reside in the low-gap film.

\subsubsection{Kinetic equation for the dynamics of quasiparticles}
In our treatment of the QP dynamics, we assume that the QP distribution functions in each pad are spatially uniform. This assumption is valid because the QP diffusion across the transmon pad is much faster than the other relevant timescales in the problem.
We parametrize the distribution functions by $x_\mathrm{qp}^L$ ($x_\mathrm{qp}^R$), the ratio between the number of QPs in the left (right) pad to the number of Cooper pairs in the \textit{low-gap film} in the same pad. We describe the dynamics of the system by the following system of equations \cite{diamond_distinguishing_2022}:
\begin{equation}
\label{eq:qp-kin}
\begin{array}{c}
\frac{d}{dt}x_{\mathrm{qp}}^{L}=
g_{L}-s_{L}x_{\mathrm{qp}}^{L}-r_{L}(x_{\mathrm{qp}}^{L})^{2}+\gamma_{R\rightarrow L}x_{\mathrm{qp}}^{R}-\gamma_{L\rightarrow R}x_{\mathrm{qp}}^{L}\\
\frac{d}{dt}x_{\mathrm{qp}}^{R}=
g_{R}-s_{R}x_{\mathrm{qp}}^{L}-r_{R}(x_{\mathrm{qp}}^{R})^{2}-\gamma_{R\rightarrow L}x_{\mathrm{qp}}^{R}+\gamma_{L\rightarrow R}x_{\mathrm{qp}}^{L}
\end{array}
\end{equation}
Terms $g_L$ and $g_R$ describe the generation of the QPs. The exact origin of the QPs is unknown, they can be generated by stray radiation or cosmic rays. The second term in the right hand side of Eq.~\eqref{eq:qp-kin} describes the relaxation of the QP concentration. {This term arises, for example, due to QP being trapped away from the junction} by the normal cores of superconducting vortices \cite{wang_measurement_2014, savich_quasiparticle_2017}. The vortices can be present due to incomplete screening of the external magnetic field by the magnetic shield enclosing our device. The third term in Eq.~\eqref{eq:qp-kin} describes the effect of QP recombination -- two QPs might come together, disappear, and emit a phonon. The final two terms in both lines of Eq.~\eqref{eq:qp-kin} describe the redistribution of the QP concentration between the transmon pads due to the QP tunneling across the junction; terms proportional to $\gamma_{R\rightarrow L}$ correspond to the tunnelling from the right pad to the left pad and vice versa for $\gamma_{L\rightarrow R}$.

Next, we link rates $\gamma_{R\rightarrow L}$ and $\gamma_{L\rightarrow R}$ to the QP tunneling rates of Section~\ref{sec:qpt}. The result is
\begin{equation}
\label{eq:trltlr}
\Gamma^{R\rightarrow L} =N_\mathrm{qp}^R \cdot \gamma_{R\rightarrow L},\quad \Gamma^{L\rightarrow R} = N_\mathrm{qp}^L \cdot \gamma_{L\rightarrow R},
\end{equation}
where $N_\mathrm{qp}^{R/L} = \Delta\cdot2\nu_{0}V_{\Delta}\cdot x_{\mathrm{qp}}^{R/L}$ is the total number of QPs in the right/left transmon pad ($V_{\Delta}$ is the total volume of the low-gap region in a single pad, {$\nu_0$ is the normal-state density of states at the Fermi level per spin projection}) and 
\begin{equation}
\label{eq:grllr}
\Gamma^{R/L\rightarrow L/R}=(1-p_1)\Gamma_{0}^{R/L\rightarrow L/R}+p_{1}\Gamma_{1}^{R/L\rightarrow L/R}.
\end{equation}
Here, $p_1$ is the probability of finding the qubit in the excited state;
rates $\Gamma_0^{R/L\rightarrow L/R}$ and $\Gamma_1^{R/L\rightarrow L/R}$ can be expressed in terms of the partial rates in Eqs.~\eqref{eq:gammaLR} and \eqref{eq:gammaRL} as
\begin{equation}
    \Gamma_0^{R/L\rightarrow L/R} = \Gamma_{00}^{R/L\rightarrow L/R} + \Gamma_{01}^{R/L\rightarrow L/R},\quad \Gamma_1^{R/L\rightarrow L/R} = \Gamma_{10}^{R/L\rightarrow L/R} + \Gamma_{11}^{R/L\rightarrow L/R} + \Gamma_{12}^{R/L\rightarrow L/R}.
\end{equation}
We stress that $\gamma_{R\rightarrow L}$ and $\gamma_{L\rightarrow R}$ are the tunneling rates \textit{per one QP}. We note that, as a result of Eq.~\eqref{eq:grllr}, the rates $\gamma^{R \rightarrow L}$ and $\gamma^{L\rightarrow R}$ \textit{linearly} depend on the excited state population of the qubit. In other words, they can be expressed as
\begin{equation}
    \label{eq:ts}
    \gamma_{R\rightarrow L} = (1-p_1)\gamma_{R\rightarrow L}^{(0)} + p_1 \gamma_{R\rightarrow L}^{(1)},\quad
    \gamma_{L\rightarrow R} = (1-p_1)\gamma_{L\rightarrow R}^{(0)} + p_1 \gamma_{L\rightarrow R}^{(1)},
\end{equation}
where $\gamma_{R\rightarrow L}^{(0)}$ denotes the QP tunneling rate per QP for  qubit in the ground state while $\gamma_{R\rightarrow L}^{(1)}$ denotes this rate for qubit in the excited state (and similarly for $\gamma_{L\rightarrow R}$).
\subsubsection{Parity switching rate as a function of qubit excited state population}
Now, we calculate the dependence of the total parity switching rate on the population of the excited state of the qubit. As announced before, due to the non-equilibrium kinetic effects this dependence might in certain cases be non-linear (in contrast to what was observed in the main text). We show when such kinetic effects can be neglected so that the linear dependence is recovered.

Throughout this section, we consider an experimentally relevant situation in which the transmon pads are symmetric. This allows us to assume that in Eq.~\eqref{eq:qp-kin} $g_L = g_R = g$, $s_L = s_R = s$, and $r_L = r_R = r$. We also neglect the recombination term in the kinteic equation since at the quasiparticle densities we have in our experiment\footnote{According to the literature $r \sim 10^7\:\mathrm{sec}^{-1}$ \cite{wang_measurement_2014} and $s$ is in the range from $10\:\mathrm{sec}^{-1}$ \cite{diamond_distinguishing_2022} to $10^3\:\mathrm{sec}^{-1}$ \cite{wang_measurement_2014} for systems similar to ours. Since in our case the quasiparticle density is low, $x_\mathrm{qp} \lesssim 10^{-9}$, the recombination can indeed be neglected.} $r x_\mathrm{qp} \ll s$. This allows us to explicitly find the steady state of Eq.~\eqref{eq:qp-kin}:
\begin{equation}
\label{eq:steady}
x_{\mathrm{qp}}^{R/L}=\frac{g}{s}\left(1\pm\frac{\gamma_{L\rightarrow R}-\gamma_{R\rightarrow L}}{s+\gamma_{L\rightarrow R}+\gamma_{R\rightarrow L}}\right).
\end{equation}
If the qubit is in thermal equilibrium, $p_1 / p_0 = e^{-h f_{10}/k_B T}$, we find $x_\mathrm{qp}^L = x_\mathrm{qp}^R$ since in such a quasi-equilibrium situation $\gamma_{L\rightarrow R} = \gamma_{R\rightarrow L}$. The exact value of $x_\mathrm{qp}=g/s$ is determined by the balance between the QP generation and relaxation.

When the qubit state is out of equilibrium we generally find $\gamma_{L\rightarrow R} \neq \gamma_{R\rightarrow L}$ and thus $x_\mathrm{qp}^R \neq x_\mathrm{qp}^L$. However, as follows from Eq.~\eqref{eq:steady} as long as $\gamma_{L\rightarrow R}, \gamma_{R\rightarrow L} \ll s$ the difference between $x_\mathrm{qp}^{R}$ and $x_\mathrm{qp}^{L}$ is small and can thus be neglected. Using the fit parameters for the theory plotted in Figure 3 of the main text we obtain $\gamma_{R\rightarrow L}, \gamma_{L\rightarrow R}\lesssim 1.0\:\mathrm{sec}^{-1}$ for all possible qubit excited state populations {(the number is small due to the large volume of transmon pads)}. Therefore, if we take the relaxation rate to be $s\gtrsim 10\:\mathrm{sec}^{-1}$ -- which is based on \cite{wang_measurement_2014, diamond_distinguishing_2022} -- the redistribution of the QP density between the transmon pads should be negligible, validating the theory used in the main text of the manuscript.

\begin{figure}[h]
\includegraphics[scale=1.0]{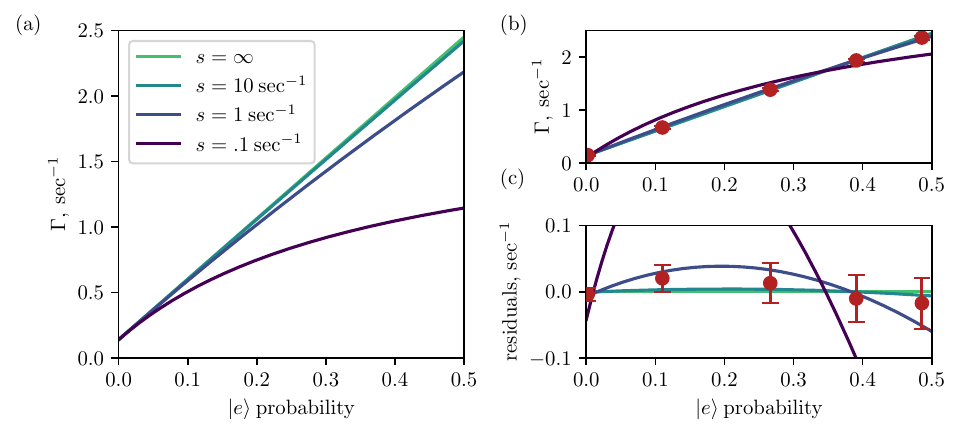}
\caption{
(a) Expected parity-switching rate as a function of qubit excited state population for different trapping rates [see Eq.~\eqref{eq:qp-kin}] with a fixed $x_\mathrm{qp}$ and $\delta\Delta$. The total rate is obtained by adding a small constant contribution $\Gamma_\mathrm{offset}$ to the QP tunneling rate given by Eq.~\eqref{eq:qpt-vs-trap} ($\Gamma_\mathrm{offset}$ accounts for the effect of absorption of high-frequency photons). There is appreciable curvature when the trapping rate is sufficiently low. This happens due to redistribution of the QPs between the two pads that can occur when the qubit is out of equilibrium. (b) Fits of the data at $20 \: \mathrm{mK}$ with different values of trapping rate. Here, $x_{qp}$ and $\Gamma_\mathrm{offset}$ are the free parameters for each curve. (c) Deviation from a linear fit. The curvature of the data is sufficiently small to allow us to put a lower bound on the trapping rate of about $1 \: \mathrm{sec^{-1}}$. There does appear to be some non-zero curvature, suggesting a measurably small trapping rate of $3 \: \mathrm{sec^{-1}}$, but the curvature is not statistically significant and it is not consistently present across the data at different temperatures.}
\label{fig:tunneling-vs-trapping}
\end{figure}

While the comparison between the tunneling rate in our setup and the \textit{expected} QP relaxation rate is promising (as it allows us to neglect the kinetic effects in the main text), we do not have a \textit{direct} measurement of the relaxation rate in our system. We note, however, that if the relaxation rate $s$ would be comparable to the tunneling rates $\gamma_{R\rightarrow L}$ and $\gamma_{L\rightarrow R}$, the dependence of the total parity switching rate $\Gamma$ on the excited qubit state population would turn out non-linear, in sharp contrast to our observations (see Figure~3(a) of the main text). To see this we calculate the total parity switching rate due to the QP tunneling as 
\begin{equation}
\label{eq:qpt-vs-trap}
\Gamma^\mathrm{qp} =(1-p_1)\left(\Gamma_{0}^{R\rightarrow L}+\Gamma_{0}^{L\rightarrow R}\right)+p_{1}\left(\Gamma_{1}^{R\rightarrow L}+\Gamma_{1}^{L\rightarrow R}\right)
\end{equation}
where we assume that $x_\mathrm{qp}$ is different in the two transmon pads as given by Eq.~\eqref{eq:steady}. This results in
\begin{equation}
\label{eq:non-lin}
\frac{\Gamma^\mathrm{qp}}{2\Delta\nu_{0}V_{\Delta}}=\frac{g}{s}\left(\gamma_{R\rightarrow L}+\gamma_{L\rightarrow R}-\frac{(\gamma_{L\rightarrow R}-\gamma_{R\rightarrow L})^{2}}{s+\gamma_{L\rightarrow R}+\gamma_{R\rightarrow L}}\right).
\end{equation}
The first two terms in the bracket in Eq.~\eqref{eq:non-lin} supplemented with Eq.~\eqref{eq:ts} describe the linear dependence of the parity switching rate on the excited state population $p_1$. The third term, however, results in a deviation from the linear trend. Such a non-linearity becomes pronounced as $\gamma_{L\rightarrow R} + \gamma_{R\rightarrow L}$ approaches $s$. We demonstrate this behaviour in Figure~\ref{fig:tunneling-vs-trapping} where plot the dependence of the parity-switching rate $\Gamma$ on the excited state population $p_1$ for different relaxation rates $s$ (assuming that the total number of QPs is fixed which we achieve by varying the generation rate $g$).

Since we do not see the non-linearity in the dependence of $\Gamma$ on $p_1$, we believe that the assumption of QPs at the two sides of the junction being in equilibrium with each other is well justified.

\subsection{Alternative interpretation of thermal activation}
Activation behavior of the QP tunneling rates might have an interpretation different from superconducting gap difference at the Josephson junction. Namely, the activation might be observed in the presence of subgap levels in the pads even if the (averaged) gaps are similar across the junction. Indeed, at small temperatures subgap levels would trap the QPs preventing their tunneling across the junction. At higher temperatures the QPs would delocalize from the levels. Assuming the trap depth has a certain characteristic energy scale, this would result in the activation of the parity switching. Activation of the parity switching rate due to the presence of (intentional) QP traps was observed in \cite{pan_engineering_2022}.

We believe, however, that this interpretation is incompatible with our data. Indeed, any model in which the gap difference at the junction is absent would predict a particular ratio between the parity switching rate in the excited and in the ground state \cite{glazman_bogoliubov_2021}
\begin{equation}
    \label{eq:ratio}
    \frac{\Gamma_1}{\Gamma_0} \sim \sqrt{\frac{\pi\Delta^2}{h f_q k_B T}} \sqrt{\frac{E_C}{8E_J}}.
\end{equation}
This result does not depend on the presence of the traps in the pads. The traps effectively modify the QP concentration 
$x_\mathrm{qp}$ making it temperature dependent. However, when computing the ratio of rates, the concentration cancels. At $T = 20\:\mathrm{mK}$ equation~\eqref{eq:ratio} results in $\Gamma_1 / \Gamma_0 \sim 6$ for the parameters of our experiment. In our data, however, the ratio between the rates is much higher, $\Gamma_1 / \Gamma_0 \sim 34$, invalidating this model. In contrast, as we showed, large ratio between $\Gamma_1$ and $\Gamma_0$ is a hallmark feature of the model with a gap difference at the junction.

\subsection{Pair-breaking by stray high-frequency photons}
\label{sec:pat}
In our system, the parity switching rate in the ground state of the qubit saturates at small temperature, see Figure~3(b) of the main text. At the same time, the parity switching due to the QPs should reduce indefinitely with temperature.
We attribute the saturation of the ground state rate to the presence of a constant contribution to the parity switching which stems from pair-breaking by stray photons with frequency $h f > 2 \Delta \sim 100\:\mathrm{GHz}$. In this section, we discuss in detail this parity-switching process. Using the theory presented in \cite{houzet_photon-assisted_2019}, we estimate the contribution of the photon-induced pair-breaking to the parity switching in the excited transmon state and show that this rate is close to rate of pair-breaking in the ground state. The fact that the total observed parity switching rate in the excited state is much higher than that in the ground state allows us to attribute $\sim 95\%$ of excited-state parity switching to the QP tunneling. For this discussion, we neglect the presence of gap difference at the Josephson junction as it is not consequential for the absorption of photons with $f > 2\Delta \gg \delta \Delta$.

We start by reviewing the photon-induced pair-breaking process. We assume that the transmon is coupled to electromagnetic environment with a non-zero density of modes at frequencies $f \sim 2\Delta$. The photons residing in these modes can be absorbed by the transmon. Such an absorption results in the breaking of a Cooper pair into two quasiparticles, one at each side of the Josephson junction. According to \cite{houzet_photon-assisted_2019}, the rate of pair-breaking absorption is given by 
\begin{gather}
\label{eq:gamma-pat}
\Gamma_{if}^{\mathrm{ph}} = S_-^\mathrm{ph}(f_{fi})|\langle f|\cos\frac{\varphi}{2}|i\rangle|^2 + S_+^\mathrm{ph}(f_{fi}) |\langle f|\sin\frac{\varphi}{2}|i\rangle|^2,
\end{gather}
where
\begin{equation}
    \label{eq:sf-photon}
    S_\pm^\mathrm{ph}(f_{fi}) = \int d\varepsilon g(\varepsilon) \mathcal{F}^\mathrm{ph}(\varepsilon)\int_{\Delta}^\infty d\epsilon \int_{\Delta}^\infty d\epsilon^\prime \frac{\epsilon\epsilon^\prime \pm \Delta^2}{\sqrt{\epsilon^2 - \Delta^2}\sqrt{{\epsilon^\prime}^2 - \Delta^2}}\delta(\epsilon + \epsilon^\prime + h f_{fi} - \varepsilon).
\end{equation}
Here, $\mathcal{F}^\mathrm{ph}(\varepsilon)$ is the distribution function of the high-energy photons incident on the transmon and $g(\varepsilon)$ is a coupling parameter which depends on the electromagnetic environment.

To proceed, we need to specify the distribution function $\mathcal{F}^\mathrm{ph}(\varepsilon)$. To this end, we note that as long as the radiation shield (see Fig.~2 of the main text) is sufficiently good, we can assume that the high-energy photons reach the sample by passing through the Eccosorb filter. Before reaching the filter, the radiation most likely has temperature $\gtrsim 2\Delta / k_B$ since otherwise it would not contribute to the parity switching; thus, the energy distribution of the photons is broad before filtering. The filter multiplies this broadband distribution by a sharp cutoff factor $\exp(- f / f_0)$ with $f_0 \sim 10\:\mathrm{GHz}$ based on the length of the filter \cite{halpern_far_1986}. Since $h f_0 \ll 2\Delta$, we can then take
\begin{equation}
\mathcal{F}^\mathrm{ph}(h f) = \mathcal{F}^\mathrm{ph}(2\Delta)\exp\left(\frac{h f - 2\Delta}{h f_0}\right).
\end{equation}
The assumption of a sharp exponential cutoff allows us to neglect the energy dependence of $g(\varepsilon)$ in integrals in Eq.~\eqref{eq:sf-photon}.
In this case the integrals can be explicitly calculated. This results in 
\begin{equation}
        S_+^\mathrm{ph}(f_{fi}) = g_0 \exp\left({-\frac{f_{fi}}{f_0}}\right),\quad
        S_-^\mathrm{ph}(f_{fi}) = g_0 \exp\left({-\frac{f_{fi}}{f_0}}\right)\frac{h f_0}{2\Delta}.
\end{equation}
where $g_0 = \pi \Delta g(2\Delta)\mathcal{F}^\mathrm{ph}(2\Delta)h f_0$. Using this formula we can compute the total rate of the parity switching in the excited state, $\Gamma_1^\mathrm{ph}$, and in the ground state, $\Gamma_0^\mathrm{ph}$. To this end, we relate the total rates to the partial rates, $\Gamma_0^\mathrm{ph} = \Gamma_{00}^\mathrm{ph} + \Gamma_{01}^\mathrm{ph}$ and $\Gamma_1^\mathrm{ph} = \Gamma_{10}^\mathrm{ph} + \Gamma_{11}^\mathrm{ph} + \Gamma_{12}^\mathrm{ph}$. Calculating the matrix elements of $\cos \frac{\varphi}{2}$ and $\sin \frac{\varphi}{2}$ in the transmon limit, we obtain
\begin{equation}
    \label{eq:pat-estimate}
    \frac{\Gamma_1^\mathrm{ph}}{\Gamma_0^\mathrm{ph}} = \frac{\frac{h f_0}{\Delta} + \sqrt{\frac{E_C}{2E_J}}\exp\left(-\frac{f_q}{f_0}\right) + \sqrt{\frac{E_C}{2E_J}}\exp\left(\frac{f_q}{f_0}\right)}{\frac{h f_0}{\Delta}  + \sqrt{\frac{E_C}{2E_J}}\exp\left(-\frac{f_q}{f_0}\right)}\sim 2.2,
\end{equation}
where the last numerical estimate is obtained for the parameters of our experiment. Based on Eq.~\eqref{eq:pat-estimate}, we estimate $\Gamma_1^\mathrm{ph}\sim 0.26\:\mathrm{sec}^{-1}$ which is only about $5\%$ of the total observed rate $\Gamma_1$ at the base temperature (and less than $5\%$ at higher temperatures). Vast majority of the parity switching in the excited state can thus be attributed to the tunneling of QPs. This consideration justifies the use of a similar photon-assisted parity switching rate for the two qubit states in Figure 3 of the main text.

\bibliography{references.bib}